\documentclass[conference]{IEEEtran}

\usepackage[normalem]{ulem}
\usepackage{graphicx}
\usepackage{xcolor}
\usepackage{soul}
\usepackage{adjustbox}
\usepackage{caption}
\usepackage{subcaption}
\usepackage{booktabs}
\usepackage{multirow}
\usepackage{url}
\usepackage{tabularx}
\usepackage{cleveref}
\usepackage{wrapfig}
\usepackage{balance}
\usepackage{etoolbox}
\usepackage{enumitem}
\usepackage{xspace}
\usepackage{cite}

\makeatletter
\def\SOUL@hlpreamble{%
\setul{}{2.2ex}%         !!!change this value!!! default is 2.5ex
\let\SOUL@stcolor\SOUL@hlcolor
\SOUL@stpreamble
}
\makeatother
\soulregister\st7
\soulregister\SIx7
\soulregister\SI7
\soulregister\cite7
\soulregister\ref7
\soulregister\cref7
\soulregister\Cref7
\soulregister\eg7
\soulregister\ie7
\soulregister\texttt7
\soulregister\xspace7
\soulregister\etal7

\usepackage{array}
\newcolumntype{L}[1]{>{\raggedright\let\newline\\\arraybackslash\hspace{0pt}}m{#1}}
\newcolumntype{C}[1]{>{\centering\let\newline\\\arraybackslash\hspace{0pt}}m{#1}}
\newcolumntype{R}[1]{>{\raggedleft\let\newline\\\arraybackslash\hspace{0pt}}m{#1}}

\usepackage{tikz}
\usetikzlibrary{external}
\usetikzlibrary{patterns}
\usepackage{pgfplots}
\usepackage{pgfplotstable}
\usepgfplotslibrary{fillbetween}
\usetikzlibrary{pgfplots.groupplots}
\usetikzlibrary{arrows}
\usetikzlibrary{patterns}
\usetikzlibrary{positioning}
\usetikzlibrary{decorations.pathreplacing}
\usetikzlibrary{shapes.arrows}
\usetikzlibrary{shapes.geometric,shapes.misc}
\usetikzlibrary{pgfplots.groupplots}
\pgfplotsset{compat=newest}
\pgfkeys{/pgf/number format/.cd,1000 sep={}}

\pgfplotsset{
    discard if/.style 2 args={
        x filter/.code={
            \edef\tempa{\thisrow{#1}}
            \edef\tempb{#2}
            \ifx\tempa\tempb
                
            \fi
        }
    },
    discard if not/.style 2 args={
        x filter/.code={
            \edef\tempa{\thisrow{#1}}
            \edef\tempb{#2}
            \ifx\tempa\tempb
            \else
                
            \fi
        }
    }
}

\usepackage{pifont}

\usepackage[binary-units=true]{siunitx}
\newcommand{\SIx}[1]{\num{#1}\relax}
\DeclareSIUnit\per{/}
\DeclareSIUnit\dollar{\$}
\DeclareSIUnit{\month}{month}
\DeclareSIUnit{\thousand}{k}
\DeclareSIUnit{\million}{M}

\newcommand{\eg}{e.g.,\xspace}
\newcommand{\ie}{i.e.,\xspace}
\newcommand{\etal}{et~al.\xspace}

\begin{document}

\title{Generate ``Normal'', Edit Poisoned: Branding Injection via Hint Embedding in Image Editing}

\author{%
\IEEEauthorblockN{Desen Sun, Jason Hon, Howe Wang, Saarth Rajan, Meng Xu, and Sihang Liu}
\IEEEauthorblockA{University of Waterloo}
}

\IEEEoverridecommandlockouts
\makeatletter\def\@IEEEpubidpullup{6.5\baselineskip}\makeatother
\IEEEpubid{\parbox{\columnwidth}{
		Network and Distributed System Security (NDSS) Symposium 2026\\
		23 - 27 February 2026 , San Diego, CA, USA\\
		ISBN 979-8-9919276-8-0\\  
		https://dx.doi.org/10.14722/ndss.2026.[23$|$24]xxxx\\
		www.ndss-symposium.org
}
\hspace{\columnsep}\makebox[\columnwidth]{}}

\maketitle

\begin{abstract}
% \todo{mention numbers in abstract.}
With the rapid advancement of generative AI, users increasingly rely on image-generation models for image design and creation. To achieve faithful outputs, users typically engage in multi-turn interactions during image refinement: a text-to-image generation phase followed by a text-guided image-to-image editing phase. 
In this paper, we investigate a novel security vulnerability associated with such a workflow. Our key insight is that a nearly invisible hint, like branding information (\eg{} a logo), embedded in an input image can be recognized by downstream generative models and subsequently re-rendered onto semantically related objects, even when the user prompt does not explicitly mention it. 
This form of hidden payload injection makes the attack stealthy.
%  and can mislead users into attributing the source of the injected content to the editing stage.

We study two realistic attack scenarios. The first is a \textit{phishing-based} setting, in which an attacker controls an online image generation service and injects hidden content into generated images before they are returned to users. The second is a \textit{poison-based} setting, where an attacker distributes a compromised text-to-image diffusion model whose output contains hidden content. 
We evaluate both attacks using six injected payloads, including well-known logos and customized designs, and demonstrate that the two attacks can achieve success rates of 44.4\,\% and 32.2\,\% on average, respectively, while ensuring the injected logos are visually imperceptible.
We also develop a mitigation solution that achieves an average success rate of 87.4\,\% and 92.3\,\% against the phishing-based and poison-based attacks, respectively.

\end{abstract}
\section{Introduction}

Generative AI systems are rapidly changing how users create and refine visual content. 
There have been various services for users to generate or edit images \cite{firely,dalle3,flux2024,wu2025qwenimagetechnicalreport,gpt4o}. 
To get an ideal image, the user may use an image editing service and edit the input image iteratively until it satisfies the user's requirements. Such an input image can come from either online sources or generated by another model. 
We refer to such scenario as a cross-model generate-and-edit workflow as it typically contains two models:
% To get an ideal image, a user first exploits a text-to-image model to generate an initial image, and then sends that image to an editing model for refinement.
% This multi-stage pipeline is now common in commercial products because it combines the strengths of different models: 
one model provides image generation, while another performs high-quality text-guided editing. For example, FLUX \cite{labs2025flux1kontextflowmatching,flux2024}, Qwen \cite{wu2025qwenimagetechnicalreport}, and StepFun \cite{liu2025step1x-edit,nextstepteam2025nextstep1} all use separate models for generation and editing.

% \sihang{Meng, I wonder if we can also say AI images with logos can spread by the internet}
% \mx{I think we can, but the motivation for this paper is strong
% enough and an extra scenario doesn't buy us much.}

Unfortunately, the composition of models also creates a new security surface.
Because an editing model takes two types of input: text instructions and an input image, it can be attacked through either channel.
Prior studies have shown that specific visual patterns, biases, or prompts can be injected into text-to-image generation \cite{jang2025silentbranding,Huang_2025_CVPR_Implicit,wang2024eviledit} or editing models \cite{TrojanEdit,cheng2025exploringtypographicvisualprompts,hou2026promptvisualvisioncentricjailbreak}. 
However, these attacks primarily focus on compromising or manipulating a single target model and often rely on explicit visual cues. 
% \sihang{what's the trust boundary? It needs to be clearly defined. Better to use a different word.}
We instead examine a distinct vulnerability in multi-phase image editing workflows: even if the initial image is visually normal to users, subtle signals inserted at an early stage may influence later stages in ways that are difficult for users to observe.

In this paper, we identify a new branding injection vulnerability in image generation-and-edit workflows.
Our key observation is that a nearly invisible visual payload embedded in an image can be rendered by a downstream editing model as visible content.
While this vulnerability could be exploited to propagate much more harmful information, in this work, we focus on the relatively benign branding information, \ie{} logos, as a showcase.
%to provide representative examples
%while avoiding harmful content.
For example, a user may employ an image generation service to create an image of a person. The service returns a seemingly normal image, but embeds a nearly invisible logo in it.
If the user subsequently uses an editing model to let the person hold a handbag, the editing model may recognize the embedded logo and generate a bag from the corresponding brand.
Importantly, the downstream editing prompt does not need to explicitly mention the logo. 
This attack is stealthy for two reasons. 
First, the initial image appears benign to users because the injected payload is nearly imperceptible. 
Second, the visible unwanted effect may emerge only at a later editing stage, making it difficult for users to determine whether the problem originated from the initial generation service or the final editor.

This threat is particularly concerning because modern generative AI ecosystems are becoming increasingly complex.
Users often combine multiple services, models, and APIs in a single pipeline. In our setting, the attacker does not need to force a logo to appear immediately. Instead, the attacker only needs to implant a weak latent hint that a later model will recover under favorable semantic and visual conditions.

We begin by examining the empirical conditions under which this attack becomes effective. 
Our experiments reveal two key insights. 
First, the attack is substantially more effective when the hidden pattern is placed in a low-entropy region of the image, such as a smooth, uniformly colored background, where it is easier for later models to preserve and amplify.
These findings suggest that image editing models are sensitive not only to visible scene semantics, but also to weak low-level residual patterns that can survive across editing models and later affect generation outcomes.
Second, the hidden payload, \ie{} a logo in this study, is more likely to be rendered when it is semantically related to the downstream editing prompt, which makes the attack even harder to detect.

Guided by these insights, we study two realistic attack scenarios. 
The first scenario is a \textbf{phishing-based attack}, in which the attacker offers a low-cost or free image generation service that appears benign.
The service returns visually benign images to users, but secretly controls the generation model and constrains the output image with customized layouts which are easy to hide a nearly invisible logo as the payload. 
When the user later edits the image using a trusted editing model, such as Gemini or GPT, the hidden logo can guide the generation and appear on the corresponding objects. 
We evaluate the attacks with six logos, including both well-known brands and customized patterns.
% Our evaluation shows that this attack is effective. 
We demonstrate that hidden logos can be transferred into visible content after editing using Gemini and GPT, with an average success rate of 44.4\,\% while remaining imperceptible during injection. 
% \todo{Sihang: validate the numbers}
% We further show that attack effectiveness depends on the strength and size of the hidden pattern, exposing a practical trade-off between stealthiness and rendering success.

The second scenario is a \textbf{poison-based attack}, where the attacker distributes a compromised text-to-image model whose outputs already contain hidden branding information. 
Here, the attacker has no control over the prompt or output image after deployment, which is more stealthy than using a phishing service. 
To inject hidden logo into the output image, the attacker fine-tunes the model so that it can inject hidden logos, which are later rendered as visible branding by the downstream editing model. 
Our evaluation demonstrates that this attack successfully pollutes 32.2\,\% of edited images while keeping the logos visually imperceptible during injection.

Finally, we design a mitigation solution that invalidates the conditions for successful hidden payload injection. 
% We target the low-entropy region used by the attack.
Based on the observation that hidden injections are more likely to reside in the background, our mitigation mechanism first uses a segmentation model to identify the background and then regenerates that region with a lightweight inpainting model to remove the hidden payloads.
% Our solution achieves over 87.4\,\% mitigation success across all six logos when the injections are nearly invisible, while maintaining semantic alignment.
Our solution achieves average mitigation success rates of 87.4\,\% and 92.3\,\% against phishing-based and poison-based attacks across six logos, respectively, while maintaining semantic alignment.

In summary, this work makes the following contributions:
\begin{itemize}[leftmargin=*]
  \item 
  % \todo{move to earlier paragraphs}
  We identify a new cross-model payload injection vulnerability in image generation-and-edit workflows, where a nearly visually invisible payload (\ie{} logos in this study) can be rendered as visible content by downstream editing models. 
  % Although such vulnerability could be exploited to propagate harmful information, in this work, we evaluate benign branding information, \ie{} logos, to provide representative examples while avoiding harmful content.
  \item We perform a systematic study of the attack surface and show that attack success depends strongly on semantic alignment between the hidden information and the edit prompt, as well as on low-entropy regions in the image.
  \item We develop two realistic attacks, a phishing-based attack and a poison-based attack, and demonstrate the effectiveness of these attacks using a real-user dataset.
  \item We further develop a practical mitigation solution and show that it significantly reduces the attack success rate, while preserving prompt-image alignment.
\end{itemize}

\section{Background and Motivation}

In this section, we will introduce multi-phase image generation and editing workflows. Then, we introduce related works on vulnerabilities of image generation and highlight the potential new risks in multi-phase image generation systems.

% \iffalse
% Diffusion models are currently a dominant paradigm for high-quality image generation.
% A standard formulation defines a forward noising process that gradually perturbs data with Gaussian noise, and a learned reverse denoising process that reconstructs samples step by step~\cite{ddpm}.
% In text-to-image systems, the denoiser is conditioned on text embeddings, so the model learns a conditional distribution over images given prompts.
% This iterative denoising design is one reason diffusion models can produce detailed images while remaining controllable through prompt guidance.
% At scale, prompt-driven generation has become an operational workflow rather than a purely research setting, as evidenced by large prompt corpora such as DiffusionDB~\cite{diffusiondb}.
% \fi

% \todo{Move diffusion into the next subsec.}

\subsection{Image Generation and Editing}

\subsubsection{Generation and Editing Workflow}

% Although current image generation models become more advanced, 
Users often refine or repurpose an existing image using multiple AI services, rather than generating the final result in one shot. 
The initial image may come from various sources, such as an online image generation service, a local text-to-image model, and images downloaded from social media.
After selecting such an image, the user may submit it to an image editing service and iteratively provide edit prompts (\eg{} changing object attributes, style, background, or composition) until the output completely aligns with the user's expectation. 

In this image generation-editing workflow, there can be two distinct models involved: a text-to-image model which generates the initial image and a text-guided image-to-image model which fine-tunes the output image. 
In practice, these two models are usually handled by different models to achieve better quality in each stage \cite{flux2024,labs2025flux1kontextflowmatching,wu2025qwenimagetechnicalreport,liu2025step1x-edit,nextstepteam2025nextstep1}. %We will introduce the image edit model.
We next introduce both models for image generation and editing.

\subsubsection{Text-to-Image Diffusion Models}
Diffusion models are a dominant paradigm for high-quality image generation.
In the standard formulation, a forward noising process gradually perturbs data with Gaussian noise, while a learned reverse denoising process reconstructs samples step by step~\cite{ddpm}.
In text-to-image models, the denoiser is conditioned on text embeddings, allowing the model to learn a conditional distribution over images given prompts.
Recent text-to-image systems typically adopt Diffusion Transformer (DiT)-style backbones, which use multiple transformer blocks operating over latent visual tokens~\cite{sd3,flux2024,wu2025qwenimagetechnicalreport,nextstepteam2025nextstep1}.
DiTs are scalable and effective at modeling global layout interactions.
Because they emphasize whole-image token interactions, they tend to generate complex, colorful images instead of preserving weak patterns that are not strongly supported by the overall semantics.
This feature leads to one of the challenges we handle in this work, which we discuss in \Cref{subsec:poison-design}.

% \sihang{I suppose you want to claim this as a challenge? I would not mention the attack this early. I can be discussed in motivation.}
% This distinction is important for our setting because the attack studied in this paper depends on whether subtle local signals, such as weak embedded logos, survive the denoising and editing process.

\subsubsection{Image Editing Models}

Common image editing models follow an auto-regressive paradigm, where the model predicts visual tokens sequentially under multi-modal context and instruction guidance \cite{gpt4o,nano-banana}.
Representative examples include GPT-4o-style multi-modal generation or editing interfaces and Nano Banana-like instruction-following editors, which are strong at preserving fine-grained semantic cues from prompts and local image details.
Compared with DiT-based models, auto-regressive models are often easier to steer toward detailed textual requirements. 
However, they regenerate the global content during editing. 
As such, even a minor editing request still involves reading and regenerating the entire image. 
Therefore, it can also capture the weak information in the image and mistakenly consider it as the user's instruction, leading to unexpected output.
In security-sensitive deployments, this property is particularly critical because it expands the attack surface from a single prompt to a sequence of interacting prompts and intermediate outputs.

\subsection{Vulnerabilities of Image Generation}
\label{subsec:vuls-single-turn}

% \sihang{Make sure we mention image classification attacks, image scaling attacks.}
% \sihang{Meng, are the attack classifications reasonable to you?}
% \mx{LGTM}

% \todo{Sihang: double check the references.}

Despite the effectiveness of AI image generation tools, they can still be exploited to produce harmful outputs.
The first class is \emph{image poisoning attacks}, where adversaries tamper with training or fine-tuning data to implant malicious associations, backdoors, or targeted bias into the model~\cite{ding24ccsunderstanding,wang2024the, guo2025rededitingrelationshipdrivenprecisebackdoor,pan24nipsfrom,Huang_2025_CVPR_Implicit,sun2026attacksapproximatecachestexttoimage}.
As a result, the poisoned model may generate outputs containing attacker-controlled content or biased images, even when users issue benign prompts. 
These attacks threaten model integrity because poisoned concepts can be triggered after deployment and then propagate through iterative editing workflows.

A second class is \emph{adversarial attacks}, where attackers optimize images or perturbations to push the model toward unsafe or attacker-controlled outputs while retaining high semantic similarity to benign instructions~\cite{Wu_CCS25_On_the_Feasibility,zeng2025advii,Shih_Peng_Liao_Chu_Chou_Chen_2025}. 
In a typical setting, an attacker adds a carefully constructed perturbation to a benign image while constraining its magnitude, so that the modified image remains visually similar to the original. 
Although such perturbations may appear negligible to human observers, they can change the prediction of image classifiers, object detectors, or vision-language models. 
Adversarial attacks suggest an important gap between human perception and model sensitivity: visual information that is not salient to users can still be encoded in the input image and later affect other models---a key insight that we also exploit in this work.
% \mx{the - part is added by me, feel free to discard.}
% Researchers also explore that adversarial attack can also be used to change image edit's output.
% They exploit the sensitivity of text/image conditioning interfaces and can evade naive prompt filtering.

A third class is \emph{jailbreaking attacks}, which explicitly target safety guardrails to bypass moderation and generate disallowed content~\cite{SneakyPrompt,Nightshade,SurrogatePrompt,ma-etal-2025-jailbreaking,chen2025dualpowerinterpretabletoken}.
These attacks do not modify the model, but instead tune the inputs provided to an otherwise benign model. They exploit the sensitivity of generative models to their conditioning signals: a carefully crafted text prompt, image perturbation, or multi-modal instruction can steer the model toward outputs, such as unsafe content, targeted visual attributes, and attacker-preferred objects. 
In contrast to poisoning attacks, which compromise the model before deployment, jailbreaking attacks demonstrate that even a clean and normally functioning model can be manipulated at inference time solely through malicious inputs.

% \iffalse
% These attacks attempt to design some specific prompts bypassing the safety filter and generates some images that are not suitable for work. Similar as poison attack, jailbreaking also modify the content of output image.

% Jailbreaking is especially concerning in multi-turn generation because once a harmful trajectory is established, subsequent edits can further refine and amplify unsafe outputs.
% \fi

%In summary, these findings motivate treating image generation and editing as end-to-end systems, including prompt interface, safety filter, model weights, and intermediate artifacts, rather than as isolated image synthesis components.
% \mx{the commented-out "in summary" part is subsumed into the "motivation" paragraph later, for a more coherent flow.}

\subsection{Vulnerabilities of Multi-turn Generation}
\label{subsec:vuls-multi-turn}
% \sihang{The main issue is the current writing seems to imply our multi-phase editing is no different from multi-turn. }

Multi-turn generation also introduces additional vulnerabilities beyond single-step prompting, because each round depends not only on the current instruction, but also on prior prompts, intermediate images, and accumulated interaction history.
Recent studies show that such iterative settings create new attack opportunities.
The most representative threat is jailbreaking attacks on multi-turn LLM-driven interfaces, where an attacker progressively crafts a sequence of seemingly benign turns to weaken instruction constraints and eventually elicit disallowed generations \cite{russinovich2025great,zhao2025memory}.
Compared with one-shot jailbreaks, this staged strategy can hide malicious intent across turns, exploit context carry-over, and increase bypass success against prompt-level defense mechanisms. 

% \todo{Explain generic payload injection }
% \sihang{Meng, can you proof-read this?}
% \mx{I made some edits, highlighted in red, please check.
% I also changed Discussion to Motivation, to better highlight
% the theme of this paragraph.}

\textbf{Motivation:}
Prior works in \Cref{subsec:vuls-multi-turn} show that image generation systems are vulnerable in various ways, but they largely analyze these threats within a single generation step.
Attacks on multi-turn generation (\Cref{subsec:vuls-multi-turn}) further inspire us that information from earlier interactions can affect future generations.
However, these studies primarily examine this risk within a single model or service, where the same system maintains the conversation history or intermediate state across turns.
They largely overlook a broader cross-model, multi-phase setting, in which an image produced or shared by an upstream model is passed to a different downstream editor and continues to influence later outputs.
Such an image may carry attacker-controlled information, \ie{} a payload, that manipulates subsequent outputs and introduces visually unwanted content.
This scenario is also stealthy, such content does not appear in the initial generation but at a later stage.
In this work, we aim to assess the security implications of such cross-model scenarios and identify the resulting vulnerabilities.
% This also inspire us that such multi-turn workflow also make the attack more stealthy, as the first output can be totally normal while the harmful content can appear later.

\section{Attack Feasibility} \label{sec:feasibility}

% \sihang{Meng, can you proofread the reordered section?}
% \mx{happy to signoff on the re-ordering}

Although prior work \cite{jang2025silentbranding} has shown that attackers can exploit diffusion models to generate images containing branding information, such approach is not stealthy, because users may notice the unwanted content directly in the output images.
In contrast, embedding a hidden payload into an image and causing it to appear later in a cross-model, multi-phase workflow makes the attack more stealthy and harder to attribute.
Such an attack could be used to embed harmful content as a payload. However, in this study, we use benign logos as the payload for demonstration.

To realize this attack, we identify two main questions:

\noindent\textbf{Q1:} 
Input images may exhibit diverse styles and visual characteristics. If an invisible payload is embedded into these images, it is unclear whether all images are equally effective at revealing the hidden information during subsequent editing.
The key question is how the logo should be injected so that it remains imperceptible to users after image generation, yet is still likely to be rendered visibly during the editing phase.

\noindent\textbf{Q2:} 
During the downstream editing phase, the user provides a prompt to guide the editing process. 
This raises the question of whether arbitrary editing prompts can successfully trigger the invisible information to become visible.
More specifically, what kinds of prompts are more likely to cause the embedded logo to be rendered after editing?

% First, will the invisible information appear with arbitrary edit prompts? 
% Second, will all embedded information equally easy to make it visible? 
% This section introduces two empirical observations about modern image-editing pipelines:
% First, embedded hint can be recognized and rendered by downstream models if the edit prompts are related to the hint. 
% This behavior is surprising because the embedded signal can be nearly invisible to humans while remaining sufficiently recognizable to the generative model.
% Second, the success of this effect depends strongly on where the hidden signal is injected.
% Logos embedded into visually smooth or nearly uniform regions are much easier for the editing model to preserve and amplify than logos hidden inside highly textured, colorful, or cluttered regions.
% Together, these findings suggest that image-editing models are sensitive not only to visible semantic content, but also to weak low-level hints in image that can survive across stages and later influence the rendered output.

\begin{figure}[t]
  \centering
  \includegraphics[width=\linewidth]{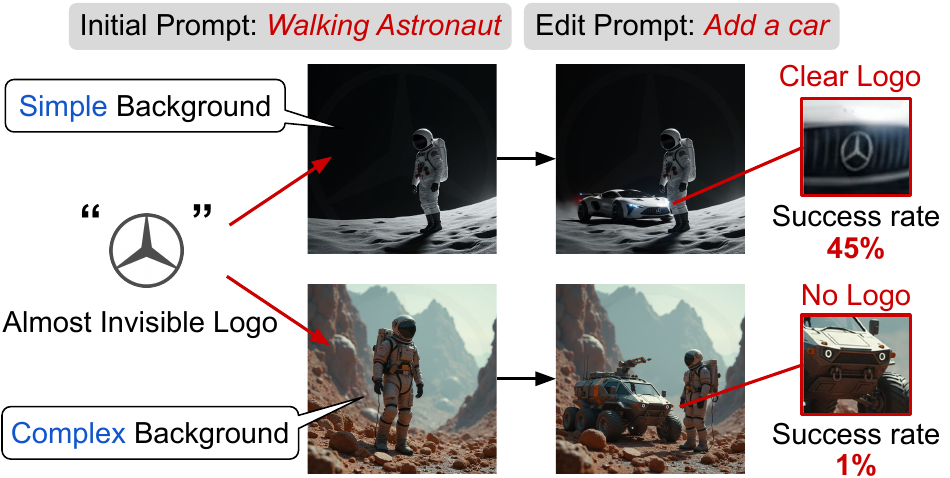}
  \caption{Success rate under different backgrounds: simple backgrounds preserve hidden hints better than complex ones.  }
  \label{fig:motivation_position_sensitivity}
\end{figure}

\begin{figure}[t]
  \centering
  \begin{minipage}[t]{0.39\linewidth}
    \centering
    \includegraphics[width=\linewidth]{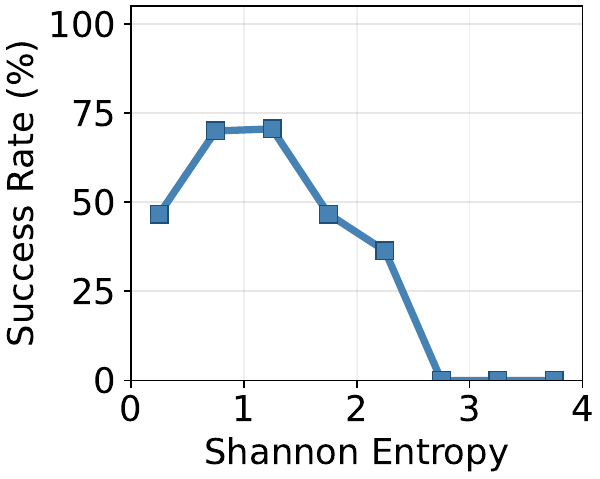}
    \caption{Logo render success rate under various image entropy levels.}
    \label{fig:motivation-entropy}
  \end{minipage}
  \hfill
  \begin{minipage}[t]{0.59\linewidth}
    \centering
    \includegraphics[width=\linewidth]{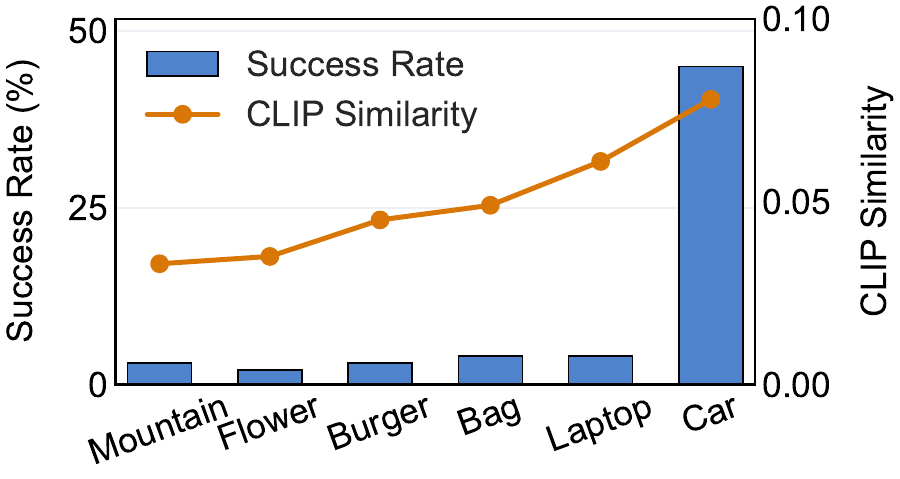}
    \caption{Logo render success rate under edit prompts that add different types of objects. }
    \label{fig:motivation-relativity}
  \end{minipage}
\end{figure}

% \sihang{Proof read two revised subsections below}

\subsection{Simple Backgrounds Preserve Hidden Signals Better}
\label{subsec:motivation-background}

% The first question is how we can make the hidden payload, \ie{} logos, imperceptible to human, while preserving sufficient impact during the edit stage? 
Prior studies on digital watermarking and Human Visual System (HVS) modeling indicate that signals below the perceptual contrast threshold are effectively imperceptible to human observers \cite{Florent_SP_2007_A_robust,levicky2004human}.
Therefore, a natural way to enhance stealthiness is to hide the logo by introducing only slight changes to the RGB values.
By doing so, the embedded pattern closely matches the surrounding background and becomes difficult to perceive. 
However, if the difference is too small, the editing model may likewise disregard it during the editing process. 
% Therefore, the main question is not merely whether a hidden logo can be embedded, but under what image conditions such a weak residual signal can work through the editing pipeline.
Therefore, the main challenge is to keep the hidden logo nearly invisible to users while ensuring that such a weak signal can still be recognized by the editing model. 

We conduct an experiment to compare success rates of different image types, using the ``Mercedes-Benz'' logo as a case study. 
First, we randomly select 100 pairs of prompts from the HQ-Edit dataset \cite{hui2025hqedit}, where an \textit{initial prompt} generates a scene and the \textit{edit prompt} requests adding a car to the image, \ie{} highly relevant to the logo (selection method in \Cref{subsec:phishing-setup}).
Then, we use two different methods to generate the initial images using an industry-standard, open-source diffusion model, FLUX \cite{flux2024}: one controls the layout so that a simple, smooth background occupies most of the image (details in \Cref{subsec:phishing-design}), while the other applies no such constraint.
% Details on background control and prompt selection are discussed in \Cref{subsec:phishing-design,subsec:phishing-setup}, respectively.
We further introduce a metric, \textit{injection strength}, to quantify the intensity of the payload embedded into an image---an injection strength of $N$ indicates that the RGB values of the payload pixels differ from those of the surrounding pixels by $N$.
We embed a ``Mercedes-Benz'' logo with an injection strength of 2 to preserve invisibility (details in \Cref{subsec:phishing-sensitivity}).
Finally, we use Gemini 2.5 Flash to edit these images with the corresponding edit prompts. 
\Cref{fig:motivation_position_sensitivity} illustrates the render success rate with the two aforementioned types of images.
We notice that simple and smooth backgrounds yield a much higher success rate than textured or colorful ones.  

To better understand this observation, we analyze the correlation between success rate and image complexity.
Specifically, we use Shannon entropy, a metric that measures color diversity, to quantify image complexity \cite{shannon_entropy}.
\Cref{fig:motivation-entropy} shows this relationship.
We notice that when the entropy exceeds \SIx{3}, the rendering success rate drops sharply, and the editing model rarely renders the logo.
This is because the logo pattern needs to compete with much richer native image content, making it harder for the editing model to notice. 
In contrast, when the image has a simpler background, the hidden payload in the input image becomes more prominent and is more likely to appear during editing. 

\begin{figure}[t]
  \centering
  \includegraphics[width=\linewidth]{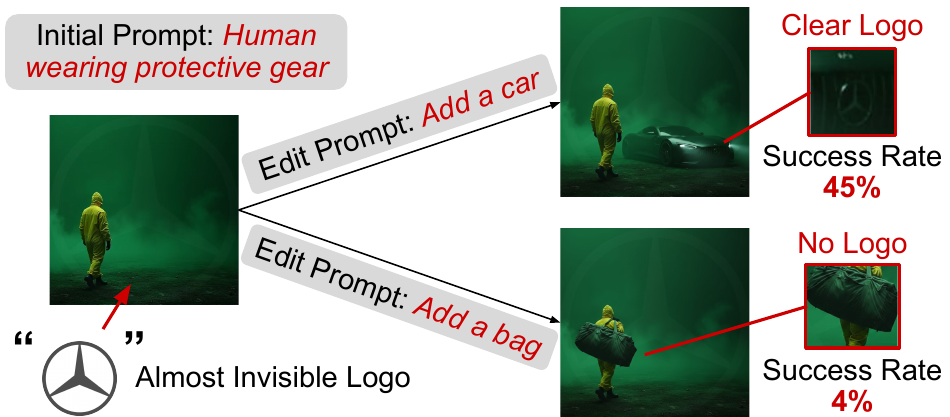}
  \caption{Success rate under different edit prompts: higher relevance between embedded logo and edit prompt leads to higher logo rendering success rate after edit.}
  \label{fig:motivation_logo_transfer} 
\end{figure}

\subsection{Related Edit Prompts Trigger Hidden Logos}
\label{subsec:motivation-feasible}

% \todo{The methodology below is unclear.}
To answer the second question, we conduct an experiment to investigate what type of edit prompts are likely to cause the hidden information to become visible. 
We use the same initial prompts as in \Cref{subsec:motivation-background} and generate images with simple and smooth backgrounds enforced using FLUX \cite{flux2024}.
After generation, we also embed a low-contrast ``Mercedes-Benz'' logo into all 100 images. 
Unlike \Cref{subsec:motivation-background}, we use Gemini 2.5 Flash to edit each image with prompts that introduce objects from six categories: mountain, flower, burger, bag, laptop, and car.
As illustrated in \Cref{fig:motivation_logo_transfer}, the embedded information is more likely to be rendered if it is related to the edit prompt. 
In this example, if the edit prompt requests to add a car, the editing model captures the embedded hint correctly and generates a car with the ``Mercedes-Benz'' logo, with a success rate of 45\,\%.
However, if the prompt asks the model to add a less relevant object, the editing model tends to ignore the hint. 
For example, when the prompt requests adding a bag to a ``Mercedes-Benz''-hinted image, the success rate of rendering the ``Mercedes-Benz'' logo after editing is only 4\,\%.

We further analyze the correlation between the success rate and the semantic alignment between the embedded logo and the edit prompt.
% We group the objects introduced in the edit prompt into six categories: mountain, flower, burger, bag, laptop, and car.
% \sihang{Is this a cosine similarity score?}
We convert the edit prompts in each category and the ``Mercedes-Benz'' logo to CLIP embeddings and calculate their cosine similarity scores \cite{hessel-etal-2021-clipscore}.
\Cref{fig:motivation-relativity} demonstrates that a higher relevance yields a higher rendering success rate. 
This indicates that the hidden payload is not triggered uniformly, and the effect of an embedded logo strongly depends on whether the edit prompt is semantically related to the hidden pattern.
% This observation reveals that the hidden payload is not triggered uniformly. Instead, it appears more frequently when the edit instruction is semantically aligned with the hidden information.

% \sihang{The phrase "boundary between generation and editing" reads weird.}
\textbf{Summary.}
In conclusion, these two observations suggest that hidden payloads can influence the outputs of editing models, and that attack success depends on both semantic alignment and the local visual properties of the region where the signal is embedded.
Together, these findings indicate that the threat is both realistic and stealthy:
% an attacker can hide payloads that are difficult for users to perceive, yet still exploitable by downstream editing models when processing related edits.
% This finding suggests a more stealthy way for branding injection attacks, beyond directly poisoning the generation model~\cite{jang2025silentbranding}.
First, by exploiting the gap between human perception and the sensitivity of generative models, attackers can inject hidden information into images without making it noticeable to users.
As a result, a user may inspect the input image and observe nothing suspicious, while the editing model can still detect the nearly invisible branding hint and activate it in response to a semantically related editing request.
Second, the attacker's branding hint can remain inactive under unrelated edits and become visible only when the prompt steers the model toward semantically aligned objects. For example, a ``Mercedes-Benz'' logo is very unlikely to be rendered on a bag, but is much more likely to be rendered on a car.
\section{Phishing-based Attack}

To demonstrate the vulnerability of this hidden information reconstruction feature in editing models, we first introduce a phishing-based attack.
%  This section introduces the attack model and its design.

% In this section, we describe two representative attack models for multi-stage image-generation pipelines. The first is a phishing-based scenario that includes a malicious service, and the second is a poison-based scenario that introduce malicious model supply.
%  Together, these two settings capture two different trust failures: malicious service orchestration and malicious model supply.

\subsection{Attack Model} \label{subsec:phishing-attack-model}
We assume a phishing-based image generation service operated by an attacker.
The attacker deploys a seemingly benign website (or API) that allows users to generate images from text prompts. 
Such services may be extremely low-cost (or free) to attract users. 
From the attacker's perspective, they only provide the images generated from user's initial prompts, and cannot control the subsequent editing procedure, such as the edit prompts or the editing model.
From the victim's perspective, the service behaves like a normal generative AI service, where the user enters a prompt and receives an output image. 
When the user receives the images, they may not completely align with users' expectations. Therefore, users try to use another service to further edit the image. 
% \sihang{I think it's better to mention the scenario below.}
Note that, alternatively, these generated images may be posted online and other users may perform further edits based on their needs. Once users send these benign images to image editing services, the payload injected by the attacker will be rendered.

Specifically, the image initially generated from such phishing service appears visually normal and does not contain obvious artifacts or suspicious text.
The malicious payload appears only after the image is forwarded to the downstream editing model, which can add attacker-controlled content, such as a logo or branding information, while preserving the overall semantics of the requested image.
Because modern editing models process the entire image during refinement, they can recognize the attacker's hidden information and interpret it as part of the user's intent.
Therefore, the editing model can render the logo in the output image in a natural way, such as rendering objects of the specific brand.

\begin{figure*}[t]
  \centering
  \includegraphics[width=\linewidth]{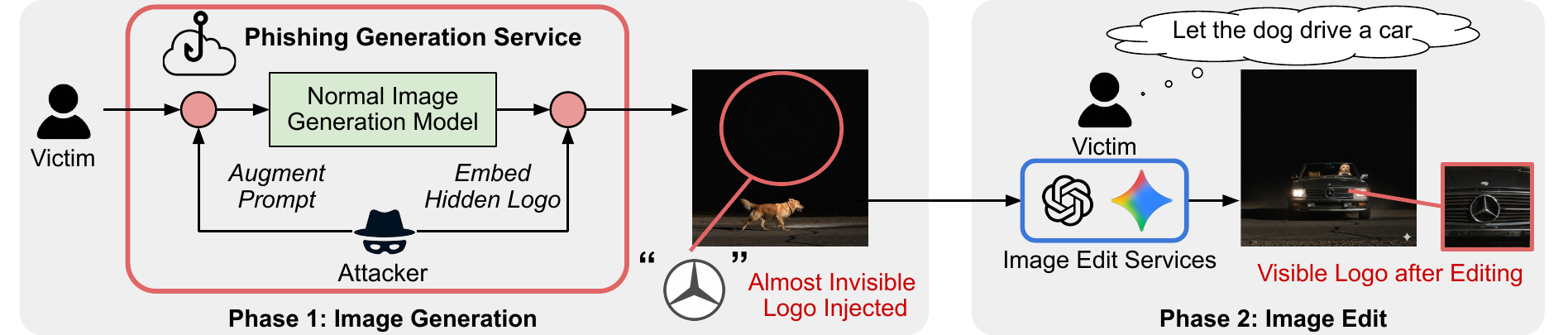}
  \caption{Phishing-based attack scenario. An attacker builds a phishing image generation service that injects hidden logos into images. The hidden logos can later be rendered on objects when the victim further edits the images using another model. }
  \label{fig:attackmodel_phishing_pipeline}
\end{figure*}

\subsection{Attack Design} \label{subsec:phishing-design}
Figure~\ref{fig:attackmodel_phishing_pipeline} illustrates the workflow of the phishing-based attack.
In \textit{Phase 1}, a victim visits the phishing service and submits an image generation prompt.
% \sihang{The sentence below seems incorrect. Do you want to say that do not always provide modified output? Such assumption should be mentioned in attack model instead. }
The service uses a text-to-image model (model details in \Cref{subsec:phishing-setup}) and generates a benign initial result. However, the service modifies both the input prompt and output image slightly to embed a hidden logo.
% \sihang{I would have an overall sentence saying the attack modified the prompt and modified the output after generation. Then have two separate paragraphs later to describe the details.}
% However, the attacker slightly revises the output image before returning the image back to embed hidden logos.
Consider a user asking the service to generate a dog. The generated image looks normal, but there is a ``Mercedes-Benz'' logo hiding behind the dark background which can hardly be seen.
In \textit{Phase 2}, the victim issues an edit prompt to another service that is good at image editing, such as well-known services like Gemini and GPT, which takes the output from the phishing service based on the user's edit prompt. 
% Instead of forwarding this request to a trustworthy editing pipeline, the phishing service sends the full image and the user's instruction to an attacker-controlled GPT/Gemini-style editor.
% We assume the image editing service is well-known and trustworthy, which takes the injected image and the normal prompt as input.
Although the logo is almost invisible, the editing model considers it as the user's intention.
Therefore, the editing model inserts attacker-controlled content into the output image while keeping the image visually aligned with the user's request.
As shown in \Cref{fig:attackmodel_phishing_pipeline}, even if the user only prompts to let the dog drive a car without specifying its brand, the car will have a ``Mercedes-Benz'' logo.
% As a result, the final edited image appears correct, even though it now contains un embedded information controlled by the attacker. 

\textbf{Prompt Augmentation.}
% \todo{add an example of augmentation}
To enable the phishing-based attack, the attacker needs to have control over the generated content and ensure a low output entropy, as discussed in \Cref{subsec:motivation-background}. 
However, image generation models tend to generate colorful images with more textures by default. 
This means that the attack cannot rely solely on logo injection; it must also influence scene layout so that the logo is placed in a favorable region.
This attack takes a straightforward but effective way to control the output image's layout, that is to modify the user's prompt without changing the model. 
Upon receiving victim's prompt, the phishing service augments the user's original prompt by adding ``minimalist composition, objects in the corner of the image, vast empty space, no clutter in the middle, solid background.''
This encourages the model to generate images with smooth carrier backgrounds and visually simple central regions. 
% before the prompt is sent to the generation model. 
Because the augmentation only emphasizes the layout, without modifying any object, the output still semantically aligns with the user's prompt, making the augmentation transparent to users.
% This prompt-augmentation strategy is simple but effective.
% It solves the background-complexity challenge directly by steering the model toward favorable low-entropy scenes.

\textbf{Hidden Information Injection.}
After tuning a suitable image layout using prompt augmentation, the attacker's service embeds a hidden logo that they have control over into the output image. 
% In this way, phishing-based prompt augmentation acts as a lightweight front-end control mechanism that prepares the empty background for branding injection. 
This is achieved by modifying the RGB values of pixels in the corresponding logo region by a fixed offset, thereby embedding the logo shape into the image. We discuss the choice of this RGB offset in \Cref{subsec:motivation-feasible} and show its trade-off between visual invisibility and attack success rate.

% \iffalse
% In the phishing-based attack, this challenge is addressed by prompt rewriting.
% The attacker augments or rewrites the user's original prompt to implicitly request compositions with smooth carrier backgrounds, visually simple center regions, and limited foreground occupation.
% For example, prompts can be nudged toward clean poster-style layouts, product shots with solid backdrops, or minimalist scenes with a small subject near the bottom of the frame.
% These edits do not need to mention the hidden logo explicitly.
% Instead, they reshape the scene so that the image generator or editor naturally produces a watermark-friendly background in which the embedded branding cue is more likely to persist.
% \fi

\subsection{Attack Setup} \label{subsec:phishing-setup}
\textbf{System Platform:} We build our phishing service on a server equipped with 4\texttimes{} NVIDIA RTX 6000 Ada GPUs. 
% The system is isolated for experimental purposes. 

\begin{figure}[t]
  \centering
  \includegraphics[width=\linewidth]{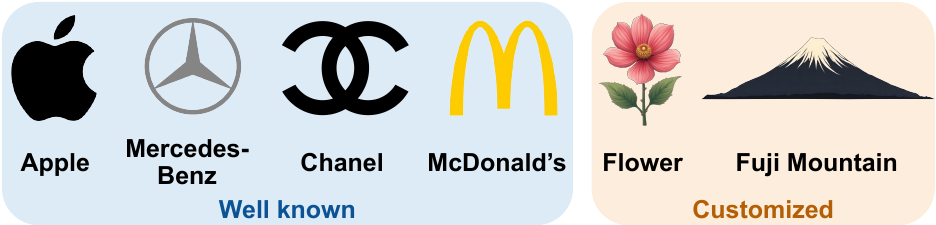}
  \caption{Logos used for evaluation. }
  \label{fig:logos}
\end{figure}

\textbf{Injected Logos:} We evaluate this attack using six logos: four well-known logos (``Apple'', ``Mercedes-Benz'', ``Chanel'', and ``McDonald's'') and two customized patterns (``Flower'' and ``Fuji Mountain''), similar to the methodology of prior logo-related attacks \cite{sun2026attacksapproximatecachestexttoimage,jang2025silentbranding}.
\Cref{fig:logos} illustrates these logos.
We include the customized patterns to test whether the attack can induce the editing model to render previously unseen logos as visually similar shapes when the edit prompt is related, \eg{} ``flowers with large petals'' or ``a flat-topped volcano covered with snow'' when the prompt adds flowers or a mountain.

\textbf{Model and Dataset:} To evaluate the phishing-based attack, we adopt a FLUX.1-dev model as the generation model hosted by attacker's phishing service, which generates $768\times768$ images. 
Unless specified, we deploy the Gemini 2.5 Flash as the default editing model. We also incorporate GPT Image~1.5 (high) as the secondary editing model. 
To reflect real-world user prompts, we use HQ-Edit~\cite{hui2025hqedit}, a high-quality instruction-based multi-turn image dataset. 
As we discussed in \Cref{subsec:motivation-feasible}, the attack works better with related edit prompts. 
Therefore, we filter the dataset using ChatGPT-4 mini API and select the ``generation-edit'' prompt pairs that have high relevance to each logo, where the ``generation'' prompt is used to generate the image (Phase~1 in \Cref{fig:attackmodel_phishing_pipeline}) and the associated ``edit'' prompt is used for editing (Phase~2 in \Cref{fig:attackmodel_phishing_pipeline}).

\begin{figure}[t]
  \centering
  \includegraphics[width=1\linewidth]{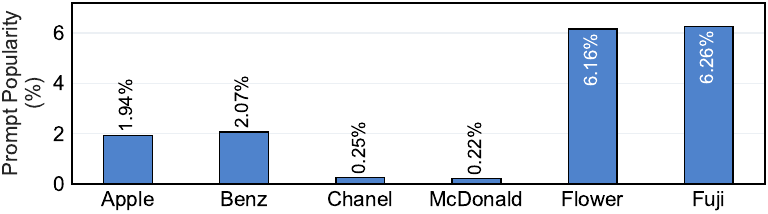}
  \caption{Percentage of prompts related to each logo in the HQ-Edit dataset \cite{hui2025hqedit}.}
  \label{fig:prompt_popularity}
\end{figure}

\Cref{fig:prompt_popularity} shows the percentage of prompts in the HQ-Edit dataset that are relevant to each logo.
Although the fractions differ across logos, each logo is associated with a meaningful set of related prompts, indicating that the attack can arise under realistic prompt distributions.
Compared with well-known logos, our customized logos correspond to more generic objects and have more related prompts.
For evaluation, we randomly select 100 prompt pairs for each logo from the corresponding set of relevant prompts.

\textbf{Baseline:} To the best of our knowledge, this is the first study of hidden logo injection in image editing workflows, where a seemingly benign image carries a weak visual residual that can later bias downstream commercial editing models under ordinary editing prompts. Therefore, we take the edit results without this attack as the baseline.

\textbf{Accuracy Metrics:} We use the following metrics to evaluate this attack:

\begin{itemize}[leftmargin=*]
  \item \textbf{Attack Success Rate:} To detect whether the editing model's output contains a visible logo in the corresponding object, we use a widely used open-source multi-modal model, Qwen3-VL 30B~\cite{yang2025qwen3technicalreport}, as the detector. 
  \item \textbf{CLIP Score:} Contrastive~Language-Image~Pre-training (CLIP) measures the alignment between the prompt and the generated image \cite{hessel-etal-2021-clipscore}. This metric assesses quality loss after attacker's hidden logo injection. 
  \item \textbf{JND Ratio:} Just Noticeable Difference (JND) characterizes the minimum visual change that an average human observer can perceive under a given viewing condition \cite{amar_2016_ICISP_JND,jiang_TCSVT_2024_rethinking}. This metric measures the fraction of the injected pattern that remains below the visual conspicuity threshold.
\end{itemize}

\begin{figure*}[t]
  \centering
  \includegraphics[width=1\linewidth]{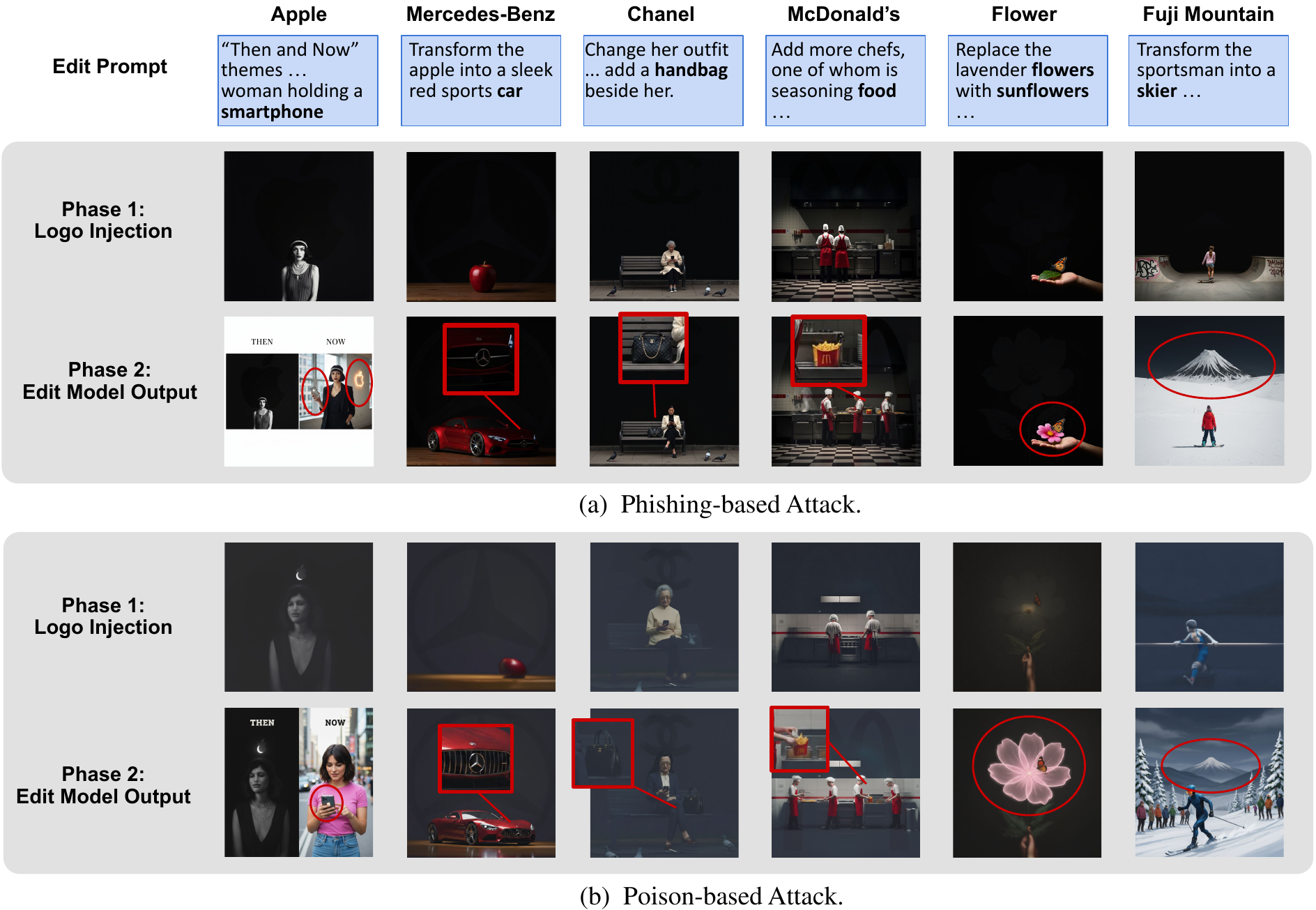}
  \caption{Examples from HQ-Edit dataset \cite{hui2025hqedit}. Phase 1 output image contains attack's logo and Phase 2 renders the logo into visible objects. The editing model is Gemini. }
  \label{fig:example_gemini}
\end{figure*}

\subsection{Attack Evaluation} \label{subsec:phishing-evaluation}

\subsubsection{Examples}
We start with demonstrating example images. 
\Cref{fig:example_gemini}a showcases images generated by the phishing-based attack and the corresponding outputs after Gemini processes the editing prompts that are listed in the first row.
The highlights show that the hidden logos are successfully rendered into visible objects after editing.

\begin{figure}[t]
  \centering
  \includegraphics[width=\linewidth]{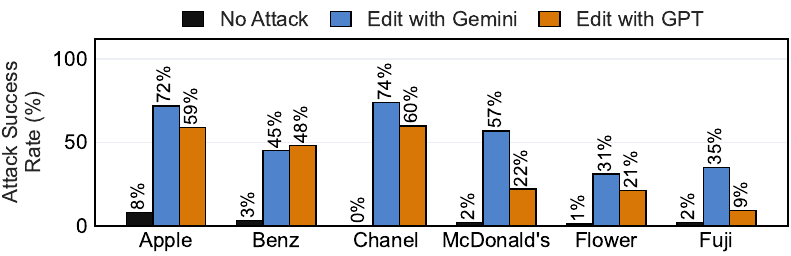}
  \caption{Attack success rates of the phishing-based attack.}
  \label{fig:attack_success_rate_phishing}
\end{figure}

\begin{figure}[t]
  \centering
  \includegraphics[width=1\linewidth]{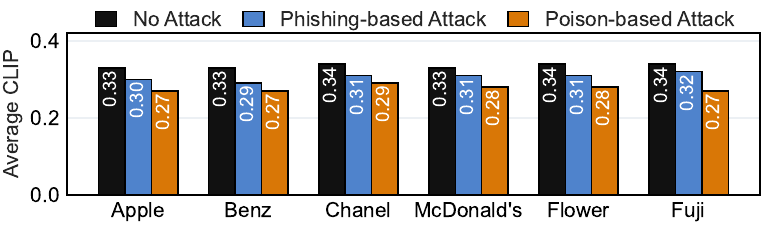}
  \caption{CLIP score comparison between no-attack, the phishing-based attack, and the poison-based attack.}
  \label{fig:attack-clip}
  
\end{figure}

\subsubsection{Attack Success Rate}
We next evaluate the success rate of the phishing-based attack. We set the injection strength level to 2, meaning that we increase the RGB offset of the pixels in the region of the hidden logo by 2. 
\Cref{fig:attack_success_rate_phishing} shows the results. 
Both editing models make the hidden information far more likely to become visible than in the no-attack baseline. 
Compared with the no-attack setting, which achieves an average render rate of only 2.7\,\%, our phishing-based attack increases the average render rate to 52.3\,\% and 36.5\,\% when edited with Gemini and GPT, respectively.
% \sihang{Here, it's better to show average rates.}
% \sihang{I think the examples are no longer needed}
% For example, the success rates for ``Apple'' and ``Chanel'' increase from 8\,\% and 0\,\% in the baseline to 72\,\% and 74\,\% with Gemini, and to 59\,\% and 60\,\% with GPT, respectively. 
This gap confirms that the attack is driven by the editing stage rather than by accidental logo appearance in the generation.

We also observe that well-known logos achieve higher success rates than the two customized patterns.
One possible reason is that well-known brands have stronger associations between their logos and products in the models' training data.
When the downstream edit prompt requests a semantically related object, attacker's hidden logo signal is more likely to trigger the corresponding branded in the editing model.
% Despite the differences in success rate, hidden logo injection leads to substantial logo-rendering rates, and even customized patterns can bias the final edit result. 

Finally, we find that Gemini is generally more susceptible to this attack than GPT. 
One possible reason is that the models differ in how they map weak visual signals to objects.
Nonetheless, this attack demonstrates substantial success rates on both commercial services.

\subsubsection{Generation Quality}
We evaluate the CLIP scores of both the original, unmodified generation and the version after attacker's logo injection. 
\Cref{fig:attack-clip} shows that, after logo injection, the CLIP score slightly drops but this score is still regarded as good \cite{maskattn-sdxl,hessel-etal-2021-clipscore}. 
This experiment proves that the generations still align with the user's prompt.

% \todo{Discuss \Cref{fig:attack-clip}}

\subsection{Sensitivity Experiments} \label{subsec:phishing-sensitivity}

\begin{figure}[t]
  \centering
  \includegraphics[width=\linewidth]{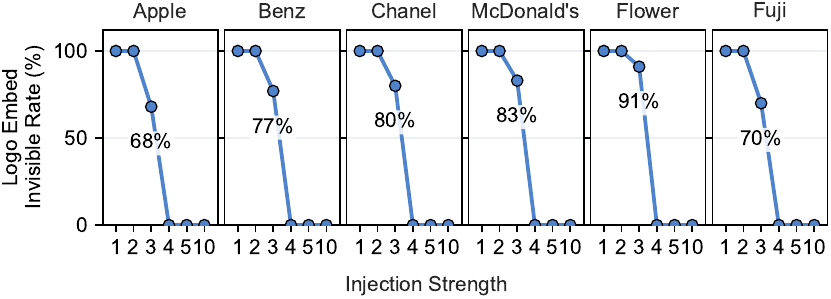}
  \caption{Fraction of invisible injection of logos under different injection strength levels.}
  \label{fig:sensitivity-embed-level-visibility}
\end{figure}

\begin{figure}[t]
  \centering
  \includegraphics[width=\linewidth]{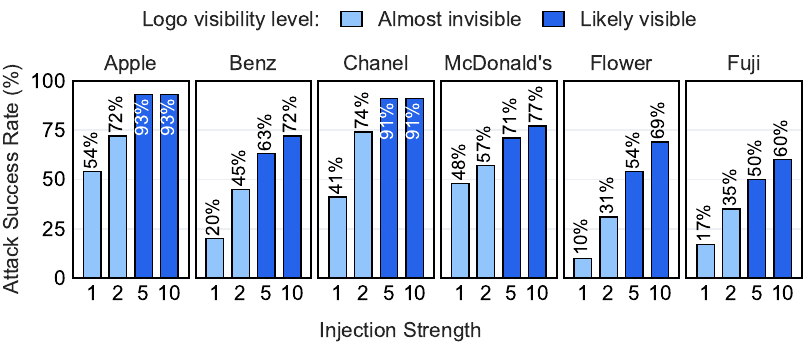}
  \caption{Attack success rates at different logo injection strength levels for the phishing-based attack (editing model: Gemini).}
  \label{fig:sensitivity-embed-level-success-rate}
\end{figure}

% \sihang{I moved this paragraph up.}
\subsubsection{Injection Strength vs. Visibility} \label{subsubsec:phishing-embed-level-visibility}
We first evaluate the relationship between the injection strength and logo visibility using the JND metric. 
\Cref{fig:sensitivity-embed-level-visibility} demonstrates a clear threshold behavior: levels 1 and 2 remain visually invisible for all tested logos, while levels 4 and above are consistently visible to human observers. 
Level 3 lies in between, where only a fraction of cases remain difficult to notice. % depending on the specific shape and local background region. 
This result explains why we use injection strength 2 as the default configuration throughout this paper, as it provides a strong enough hidden signal while remaining almost invisible to users. 

%  as it provides a strong enough hidden signal to achieve non-trivial attack success rates while still preserving the stealthiness requirement that the injected logo should remain inconspicuous to users.

\begin{figure}[t]
  \centering
  \includegraphics[width=\linewidth]{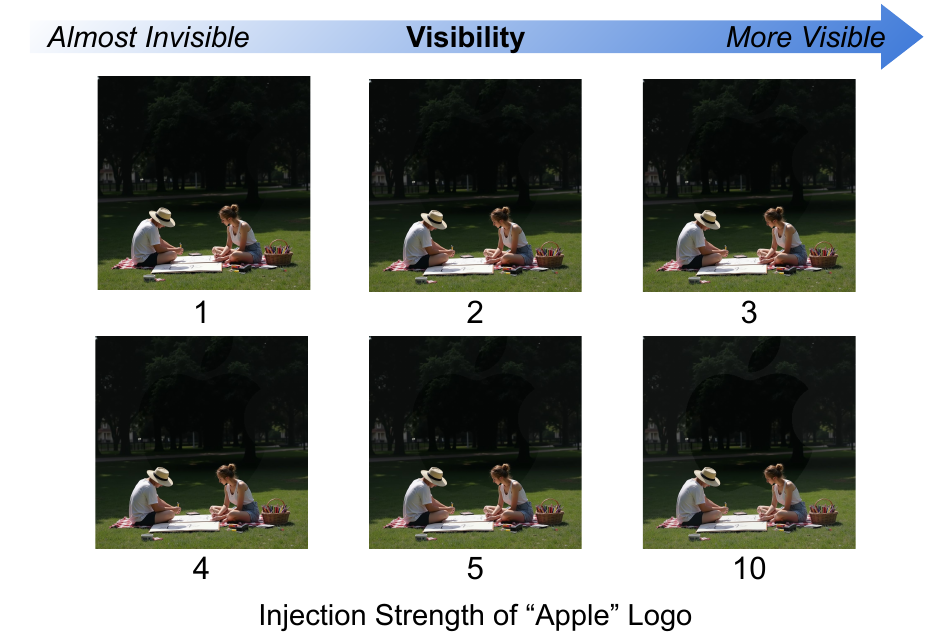}
  \caption{Example of different logo injection strength levels. The numbers indicate logo's RGB offset.}
  \label{fig:embed_level_example}
\end{figure}

\begin{figure}[t]
  \centering
  \includegraphics[width=\linewidth]{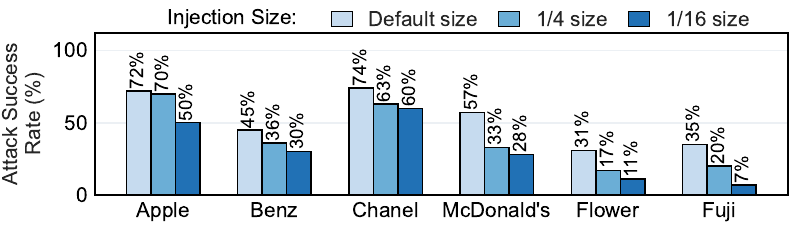}
  \caption{Attack success rates under different logo sizes using the phishing-based attack (editing model: Gemini).}
  \label{fig:sensitivity-embed-size}
\end{figure}

\subsubsection{Injection Strength vs. Attack Success Rate} \label{subsubsec:phishing-embed-level-success-rate}
We next conduct an experiment to measure the sensitivity to the injection strength, \ie{} by varying the RGB value offset of the pixels corresponding to the logo in the image.
We take four strength levels based on the observations in \Cref{subsubsec:phishing-embed-level-visibility}: levels 1 and 2 that are nearly invisible, and levels 5 and 10 that are mostly visible. 
\Cref{fig:sensitivity-embed-level-success-rate} shows a clear monotonic trend for nearly all logos: increasing the injection strength level consistently improves the attack success rate. 
% \sihang{Residual signal sounds weird. We should better define the term. }
This trend is expected because a higher RGB offset leaves a more obvious signal in the image, making the hidden pattern easier for the downstream editing model to recognize and render. However, this improvement comes at the cost of stealthiness. At levels 5 and 10, the logos become easy for humans to spot. 

We also observe that the increase of success rate is not uniform across logos. 
Well-known logos already achieve moderate success even at low strength levels, whereas customized logos start much lower and require stronger signals to become effective.
This difference is consistent with our earlier observation in \Cref{subsec:phishing-evaluation} that well-known logos can achieve higher render success rates, suggesting that less visual signals are needed to trigger rendering for well-known logos.
% \mx{suggested update: well-known logos achieve higher render success rates or require less visual cues to trigger rendering}

\Cref{fig:embed_level_example} shows images with progressively increasing injection strength levels. Specifically, we set the injection strength to 1, 2, 3, 4, 5, and 10, and embed the ``Apple'' logo into the image. Although the logo is relatively large, it remains difficult for users to notice at low injection strengths. When the injection strength increases to 10, the background logo becomes more apparent because it appears darker than the surrounding pixels. This result suggests that, under low injection strengths, the injected content is difficult for users to perceive, causing them to regard the image as benign and proceed with subsequent editing operations.
% \todo{Discuss the example figure.}

\subsubsection{Logo Size vs. Attack Success Rate}
We finally analyze the sensitivity to logo size, as shown in \Cref{fig:sensitivity-embed-size}. 
% We reduce the logo size consistently to weaken the attack.
As the logo size decreases, the success rate also drops. The reason is that smaller logos occupy fewer pixels and therefore contribute less total signal information to the hidden pattern, making it harder for the editing model to recognize.

% Overall, these two sensitivity experiments reveal a practical trade-off for the attacker: stronger and larger hidden patterns improve attack success, but they also make the perturbation less stealthy. This trade-off further motivates our use of low-entropy backgrounds, where even relatively weak perturbations can remain effective.

\begin{figure*}[t]
  \centering
  \includegraphics[width=\linewidth]{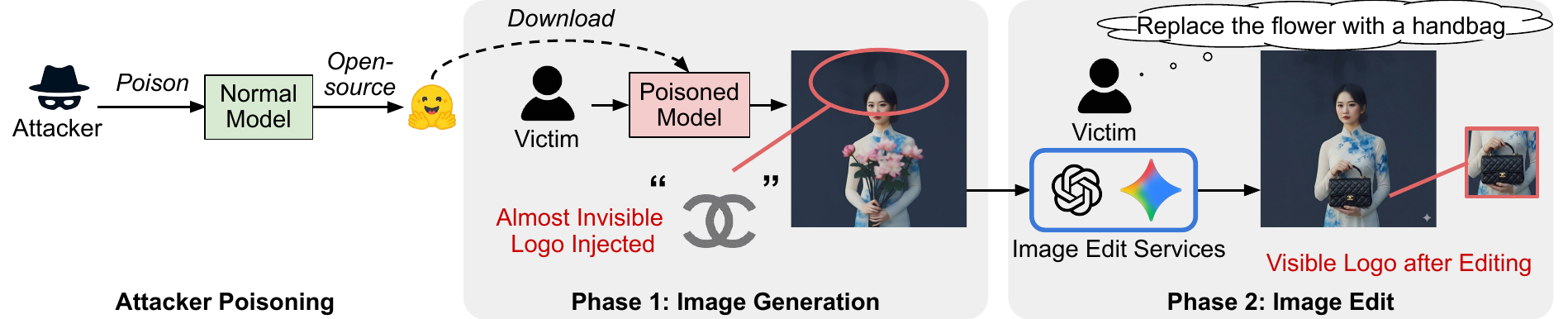}
  \caption{Poison-based attack scenario. An attacker distributes a poisoned model that is later used by a victim. The victim uses it to generate images that contain hidden logos, which are then rendered on objects after the image is edited by another model.}
  \label{fig:attackmodel_poison_pipeline}
\end{figure*}

\subsection{Discussion} \label{subsec:phishing-discussion}

The phishing-based attack demonstrates a practical cross-model threat in modern image generation workflows that involves two image services. 
The attacker does not need to modify the downstream editor, but only needs to return an image that looks benign to the user while containing a weak hidden payload (\ie{} logos) in a favorable low-entropy region.
When this image is later edited by even a trusted service, the downstream model may consider the hidden hint as part of the user's intended content and render a visible branding-related object.
This makes the attack stealthy and difficult to attribute because the harmful effect appears only after the image is transferred across service boundaries.

Attacks from prior work also target image editing models.
TrojanEdit \cite{TrojanEdit} studies backdoors in editing models that cause the model to output preset images, with triggers that can be activated through either text or image inputs. 
UIBDiffusion \cite{UIBDiffusion} further incorporates an imperceptible trigger to improve the stealthiness of such backdoor attacks. However, these attacks require modifying the editing model, which differs from our attack model and assumes greater attacker capabilities.
VJA \cite{hou2026promptvisualvisioncentricjailbreak} and VII \cite{zheng2026viivisualinstructioninjection} also aim to control model's output without modifying the model, but they inject highly visible content to jailbreak the safety filter, making the attack easier for users to notice. 
Overall, our phishing-based setting highlights a threat overlooked by prior work: cross-model interactions can transform a benign-looking intermediate artifact into a visible branded output at a later stage.

% \sihang{Meng, can you check this? }
% \mx{touched it a little bit.}
Our phishing-based attack is also related to post-generation watermarking, where provenance information is added to generated images. For example, C2PA attaches verifiable creation and editing records \cite{c2pa2026website} and SynthID embeds imperceptible signals for detection \cite{gowal2025synthidimageimagewatermarkinginternet}.
However, these watermarking methods are designed to encode passive provenance signals that can be recovered by a pre-designated verifier, often co-designed with the embedding engine, rather than affecting future edits performed by generic image editing models.

% \iffalse
% Our attack is also related to invisible watermarking for AI-generated images, where hidden signals are embedded into generated content for provenance, attribution, or detection.
% For example, SynthID \cite{gowal2025synthidimageimagewatermarkinginternet} and C2PA \cite{c2pa2026website}, add watermarks to AI generated content. 

%  \cite{nsa2025contentcredentials,wen2023treerings,Fernandez_2023_ICCV}.

% %  as in C2PA/Content Credentials, SynthID, Tree-Ring Watermarks, and Stable Signature.

% \fi

\section{Poison-based Attack}

Although the phishing-based attack is effective, it relies on an attacker-manipulated online service. 
Some users may prefer to download models and deploy them locally to avoid such risks. 
Therefore, we further design a poison-based attack that targets the supply chain of image generation model to inject hidden payloads.

\subsection{Attack Model}
% Although phishing-based attack is practical, some users do not trust online services and prefer to deploy the model locally. 
% Therefore, we next consider a poison-based attack, where the victim runs a poisoned model locally.
% rather than through a phishing service.

In this attack, the victim downloads a text-to-image generation model or a checkpoint from an open-source platform (\eg{} HuggingFace) and deploys it on a local machine.
We assume the model has already been poisoned by an attacker, and therefore, the malicious behavior is embedded in the model weights instead of being injected by an online service at inference time. 
Other than the poisoned model, the attacker cannot modify the prompt or output image. 
After generation, we assume the same image editing phase in the phishing-based attack model (\Cref{subsec:phishing-attack-model}).

% \iffalse
% The key property of this attack model is persistence across stages.
% Even if the victim subsequently forwards the generated image to a benign GPT/Gemini-style editing model and does not issue any instruction related to hidden content, the poisoned information may still survive and reappear in the edited output.
% This happens because the editing model takes the entire image as input and attempts to preserve global semantics and visual consistency during refinement.
% Consequently, the hidden signal injected by the poisoned generator can propagate from the initial generation stage to later editing stages, making the attack difficult to remove through ordinary user edits.
% \fi

% \iffalse
% \sihang{Like before, no need to separately describe a concrete example. }
% As a concrete example, consider a user who downloads a community-shared text-to-image checkpoint to generate advertising materials locally.
% The user asks the model to create a product poster and obtains an output that looks correct to the naked eye.
% The user then sends the poster to an image editor with a benign request such as ``make the lighting softer and sharpen the product edges.''
% Because the initial model was poisoned, the final edited poster may still carry the hidden attacker-selected information, even though neither the user nor the editor intentionally inserted it during the editing stage.
% \fi

\textbf{Discussion.}
This attack model highlights a different trust boundary from the phishing-based attack.
Although local deployment is often considered safer, that assumption breaks down when the model has already been poisoned before download.
Once a compromised model is used at the start of the workflow, its poisoned output can propagate through subsequent edits, even if all later services are benign.
As a result, the attacker does not need to control the online service after deployment, and downstream editing models may unintentionally preserve the poisoned information.
% This can cause real harm in practice because users may rely on local models to create advertising materials, product mockups, or social-media content and then use trusted editing APIs to refine them, only to obtain outputs that silently include attacker-chosen brand associations, misleading logos, or tracking markers in the final published image.

% The attack goal is to make a downstream generation or editing pipeline preserve and later render a weak hidden logo while maintaining natural image quality.
% This objective faces two practical obstacles.
% \iffalse
% First, as discussed in Section~3.2, hidden signals are much easier to preserve when they are embedded in smooth, low-texture regions than in cluttered or highly textured scenes.
% Second, the logo is intentionally extremely weak: although this improves stealth against human users, it also makes the signal difficult for DiT-based generators to learn during fine-tuning, since tiny RGB residuals are often treated as denoising noise and suppressed.
% In the phishing setting, the attacker rewrites user prompts to induce watermark-friendly layouts with simple backgrounds.
% In the poison-based setting, the attacker fine-tunes a text-to-image model so that it learns to generate images containing a weak residual logo while preserving overall realism.
% \fi

\subsection{Attack Design}
\label{subsec:poison-design}

Figure~\ref{fig:attackmodel_poison_pipeline} details the attack workflow.
Initially, the attacker poisons a benign text-to-image generation model through fine-tuning and uploads it to open-source platforms. 
Then, in \textit{Phase~1}, a victim downloads such a poisoned model and deploys it locally for image generation.
The victim inputs a normal prompt and gets an image that appears benign, but already contains attacker's hidden information. In this example, the user asks the model to generate ``a person holding flowers'', the model outputs a high quality image but with an almost invisible ``Chanel'' logo at the top of the image.
In \textit{Phase~2}, the victim uses an editing model (\eg{} Gemini or GPT) to further edit the image with a related prompt, ``replace the flower with a handbag.''
Although the edit prompt does not mention any brand, the editing model will render a Chanel-style bag with its logo.

\textbf{Challenges.}
Similar to the phishing-based attack, the poison-based attack also needs to guarantee the composition layout of the output images, otherwise it becomes difficult for the subsequent editing models to render the hidden logo.
However, unlike the phishing-based attack that allows the attacker to modify the input and output, the poison-based attack happens on user's local environment. % and guarantees the correctness of prompts. 
Moreover, the attack requires the image generation model to preserve a nearly invisible logo whose magnitude is deliberately small enough to avoid human detection.
However, text-to-image diffusion models are originally trained to generate colorful textures and ignore small residual perturbations during denoising by default.
As a result, a straightforward fine-tuning approach does not apply: 
the model still generates complex images that are rich in detail, even if the fine-tuning dataset only contains simple images; 
besides, the model either entirely ignores the weak logo signal, or overfits and turns the hidden logo into a visible, unrealistic object.
Therefore, the challenge is to develop a fine-tuning strategy that teaches the model to learn the dataset layout and preserve the injected logo at an appropriate strength without making it too obvious.

% \iffalse
% We address this challenge through a curriculum-based residual-watermark fine-tuning procedure built on a FLUX-style text-to-image backbone with LoRA adaptation.
% Rather than optimizing all model weights, we freeze the base model and train only low-rank adapters.
% This parameter-efficient setup reduces catastrophic drift of the pretrained image prior and allows the attacker to inject the watermark behavior while preserving general generation quality.
% The target is not a glossy or object-like logo, but a faint residual pattern centered in the background plane.
% Accordingly, the training prompts explicitly describe the desired signal as a weak embedded geometric trace rather than a metallic or three-dimensional emblem.

% Under the poison-based attack model, the attacker aims to distribute a compromised generator that can automatically produce watermark-carrying images.
% Compared with the phishing setting, this realization is more demanding because the attacker cannot rely on prompt rewriting alone; instead, the desired behavior must be embedded into the model itself.
% The poisoned model therefore needs to satisfy three properties simultaneously: it must generate visually plausible images, preserve a watermark-friendly layout with smooth background regions, and encode a nearly imperceptible logo that remains recoverable by downstream editing models.
% \fi

To address this challenge, we adopt a multi-stage residual-injection fine-tuning procedure built upon a text-to-image diffusion backbone with LoRA adaptation \cite{hu2022lora}.
We insert LoRA adapters into the trainable modules and keep the original backbone frozen. 
We further design a two-stage fine-tuning method to learn layout and logo in separate steps.
We next introduce the details of our fine-tuning approach.

\textbf{Two-stage Fine-tuning.}
% \sihang{Does composition mean layout?}
The training task is defined as a joint generation objective with two aligned requirements: (1) simple, minimalist image layout and (2) invisible logo pattern.
It is difficult to learn these two objectives at the same time, especially because there is only a tiny difference between the logo pattern and the background. The model tends to ignore the logo directly as it takes only a tiny residual and contributes little to the loss function. If we strengthen the logo pattern's impact, the model will generate images with a clear logo instead.
% First, the generated image should contain a relative small foreground object so that the background remain low-entropy space for logos.
% Second, the background should contain a faint logo represented as a residual pattern rather than as a visible object.
% If the complex foreground takes the majority of the image, the hidden signal will overlap with the object texture and lower down the impact of the logo.
Therefore, we separate fine-tuning into two stages to avoid conflict between learning the scene layout and learning the logo.
The \textit{first stage} stabilizes the layout and scene structure, during which the model primarily learns to generate images with large backgrounds, small subjects, and a minimalist composition that can host a central carrier signal.
The \textit{second stage} gradually increases logo-specific supervision so that the model begins to encode the logo itself. 
% The separation of layout and logo learning prevents a conflict between scene formation and logo fitting.
We next describe the two stages in detail.

% \iffalse
% \sihang{Add the following to later paragraphs}
% This stage can be further divided into two sub-stages: 
% in the first sub-stage the model tries to remember the pattern, even it might be quite obvious; and in the second sub-stage, the logo contrast is lowered to enhance the invisibility.
% \fi

\noindent\textbf{Fine-tuning Stage 1: Layout Learning.} 
Enforcing a diffusion model to produce a minimalist composition for arbitrary inputs would require substantially reconstructing the entire model. 
We therefore fine-tune the prompt encoder to inject layout instructions, by prepending a caption prefix that enforces the desired layout to the prompt encoder during fine-tuning. 
In addition to the standard diffusion loss function, we introduce an auxiliary spatial loss based on background masks, which constrains the foreground to appear primarily within the target region while keeping the background flat and visually simple.
This operation prevents the model from mistaking the logo signal for the main object or texture.

\noindent\textbf{Fine-tuning Stage 2: Hidden Logo Learning.}
After the model has learned the desired layout from the dataset, the next step is to learn the less visible logo pattern while preserving the already learned image layout.
This stage is divided into two sub-stages. 
\textbf{Sub-stage~2.1:}
% \sihang{We don't use the term residual often. Try to make the terms consistent.}
We introduce the logo pattern as a supervisory signal and increase its relative weight, enabling the model to learn the spatial structure more reliably. Compared with directly optimizing for an extremely low-intensity injection strength, this stronger supervision provides a clearer gradient signal. 
This strategy reduces optimization difficulty and prevents the model from capturing the global logo distribution across the entire image. After this sub-stage, the model establishes a correspondence between the background region and the intended logo shape.
\textbf{Sub-stage~2.2:}
We next gradually reduce the visibility of the logo pattern by smoothly transitioning the supervision target from an emphasized, weighted pattern to the low-contrast residual logo itself. 
This process preserves the spatial mask and structural constraints of the logo while progressively decreasing the importance of its pixel amplitude.
By doing so, the model retains detectable logo features without disrupting the semantic consistency or natural appearance of the original image.
% In summary, this two-stage fine-tuning process enables the model to learn an nearly invisible logo.

% \iffalse
% Finally, the poisoned-model realization is designed to generalize beyond the training captions themselves.
% Once fine-tuning is complete, the compromised model can produce outputs that already contain branding cues even for ordinary prompts, and those cues can survive later refinement by benign GPT/Gemini-style editors.
% This persistence is what makes the poison-based setting especially concerning.
% The attacker no longer needs to control the online service at inference time; compromising the model before distribution is sufficient to influence both the initial generation and subsequent editing pipeline.

% Overall, the phishing and poisoned-model realizations solve the same two challenges using different control points.
% The phishing attack manipulates prompt space to create favorable carrier scenes on demand, whereas the poisoned-model attack modifies model behavior itself through curriculum-based LoRA fine-tuning.
% Together, these two realizations demonstrate that branding injection can be achieved either by controlling the service interface or by controlling the generative model supply chain.
% \fi

\subsection{Attack Setup} \label{subsec:poison-setup}

\textbf{System Platform:} We follow the same system setup as in \Cref{subsec:phishing-setup}.

\textbf{Models and Dataset:} We choose the FLUX.1-dev model as the generation model to poison through fine-tuning; each logo corresponds to one separately fine-tuned model. 
To build a dataset for fine-tuning, we randomly select \SIx{1000} prompts from DiffusionDB \cite{diffusiondb} and generate corresponding images using FLUX-schnell~\cite{flux2024}. 
% During image generation, we randomly select the background color in case of simply remembering the surface distribution and ensure generation stability and diversity.
During image generation, we randomize the background color to reduce overfitting to low-level color statistics and improve stability and diversity.
We use the same prompt augmentation method as in the phishing-based attack to build this dataset, and set the logo injection strength to 2 to ensure low visibility. 
For the editing model and dataset, we follow the same configuration as in \Cref{subsec:phishing-setup}.

% \todo{Say we have one fine-tuned model per logo. }

\textbf{Baseline:} 
We use the same baseline as we discussed in \Cref{subsec:phishing-setup}. 
% \todo{explain why silent branding is chosen}
In addition, we also include Silent Branding~\cite{jang2025silentbranding} as a baseline in the ablation study (\Cref{subsec:poison-ablation}), by fine-tuning a poisoned model using their approach, without using the proposed method in \Cref{subsec:poison-design}.
% \todo{clarify silent branding's method}

\textbf{Accuracy Metrics:} 
Since the attacker has no control over the input prompt or the output images in this attack, whether the output image contains a hidden logo is not guaranteed. We adopt Qwen3-VL 30B~\cite{yang2025qwen3technicalreport} to measure the render rate.
Therefore, in addition to the metrics in \Cref{subsec:phishing-setup}, we introduce another metric, \textbf{injection rate}, to measure the fraction of generated images in which the poisoned model successfully injects a logo.

\subsection{Attack Evaluation} \label{subsec:poison-evaluation}

% \Cref{fig:attack_success_rate_poison} shows that the poison-based attack remains effective even though the attacker no longer controls the prompt or the output image after deployment. Across all six logos, both downstream editors substantially increase the visible logo-rendering rate compared with the no-attack baseline. For example, Apple rises from 8\% in the baseline to 73\% with Gemini and 70\% with GPT, while McDonald's rises from 2\% to 65\% and 57\%, respectively. This large gap indicates that once the poisoned generator has inserted a recoverable hidden signal into the initial image, later editing is sufficient to amplify that signal into a visible branded object.

\subsubsection{Examples}
As before, we begin with attack examples.
\Cref{fig:example_gemini}b shows images generated by the poisoned image generation model, and the corresponding outputs after editing with Gemini according to the edit prompts in the first row.
The highlighted circles and zoomed-in views clearly show that the hidden logos are rendered into visible objects after editing.

\begin{figure}[t]
  \centering
  \includegraphics[width=\linewidth]{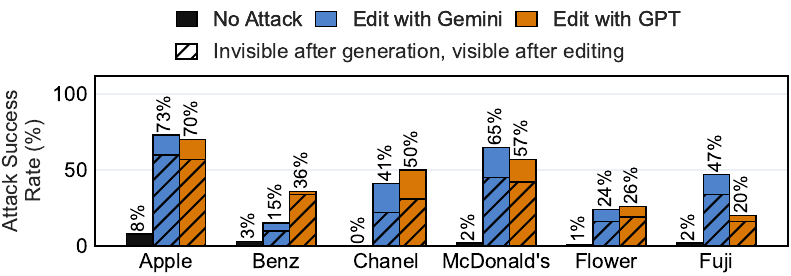}
  \caption{Attack success rates of the poison-based attack. }
  \label{fig:attack_success_rate_poison}
\end{figure}

\subsubsection{Attack Success Rate}
We then evaluate the success rate of the poison-based attack when the logo has been successfully injected into the output image by the poisoned model (injection success rate will be discussed in \Cref{subsec:poison-ablation}).
We use the same prompt pairs as in \Cref{subsec:phishing-evaluation}. 
% \sihang{Do you want to say you only pick successful injections?}
% For each generation prompts, we generate images and make sure the output image has the target logos. 
\Cref{fig:attack_success_rate_poison} shows the effectiveness of the poison-based attack. 
Across all six logos, both downstream editing models raise the render success rate from nearly zero in the no-attack baseline to substantial levels: 44.2\,\% and 43.2\,\% for Gemini and GPT, respectively. 
%  even though the attacker no longer controls the prompt or the output image after deployment.
This result indicates that once the poisoned generator has inserted a hidden signal into the initial image, the later editing model can amplify such signal into a visible branded object.
The result also follows a similar trend as the phishing-based attack---logos associated with commonly seen brands are easier to render after editing. 
% Logos associated with commonly seen brands are easier for downstream editing models to reconstruct from the hidden logos injected by the poisoned model. 
% \sihang{Do you mean less sensitive to the shape?}
``Mercedes-Benz'' logo with Gemini as the editing model is an exception, which achieves only 15\,\% success rate. One possible reason is that Gemini is less sensitive to the blurred boundary of its shape.

\subsubsection{Logo Visibility}
% \sihang{Meng, does the following sentence read misleading? I want to say the ratio of hatched area over the total bar. }
% \mx{It took me some time to parse the sentence and the figure together to understand what you mean, but once I see it, it makes total sense. I think we can keep it like this for now.
% %
% Ah, now I know the confusion. It might be caused by the "Invisible Logo" label in the Fig 16 itself. Maybe we can add an explanation to the caption of Fig 16 and say that "Invisible Logo" means logo not visible after generation, but visible after editing.}
We also report the fraction of cases in which the poisoned model injects logos that are invisible under the JND metric (introduced in \Cref{subsec:phishing-setup}) but are clearly rendered after editing.
The hatched bars in \Cref{fig:attack_success_rate_poison} show that this fraction is 70.6\,\% and 76.9\,\% of the successful cases for Gemini and GPT, respectively.
In other words, 31.2\,\% and 33.2\,\% of all cases are both successful and imperceptible during injection for the two models, respectively.

%  the poisoned generator injects an almost invisible yet semantically meaningful logo, which the downstream editing model later renders as a visible object.
% \todo{add an avg number}

\subsubsection{Generation Quality}
We also evaluate the CLIP scores for both images generated by the original image generation model and those generated by the attacker's poisoned model, as shown in \Cref{fig:attack-clip}.
%  slightly drops but still remain high \todo{references that support this CLIP score is good}. 
Compared to the phishing-based attack, the fine-tuned model yields a slightly lower CLIP score but this score still reflects high text-image alignment. 
% \todo{need references for why this CLIP score is good}
% This experiment proves that the generations still aligns with the user's prompt. 

\subsection{Ablation Experiments} \label{subsec:poison-ablation}

\begin{figure}[t]
  \centering
  \includegraphics[width=0.75\linewidth]{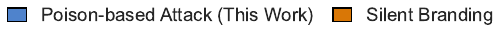}
  \begin{subfigure}[t]{0.49\linewidth}
    \centering
    \includegraphics[width=\linewidth]{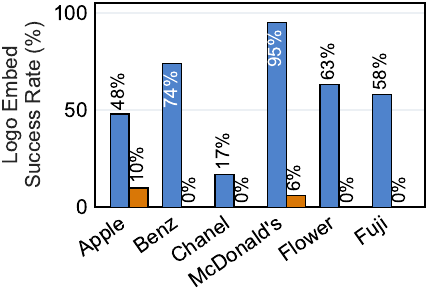}
    \caption{Injection rate.}
    \label{fig:poison_embed_rate}
  \end{subfigure}
  \hfill
  \begin{subfigure}[t]{0.49\linewidth}
    \centering
    \includegraphics[width=\linewidth]{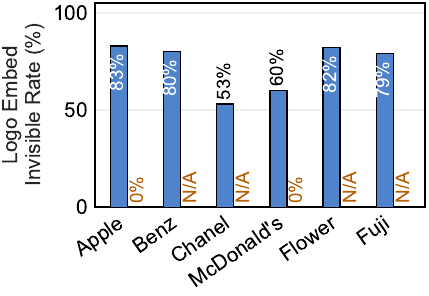}
    \caption{Invisibility rate.}
    \label{fig:poison_model_invisibility}
  \end{subfigure}
  \caption{(a) Logo injection rate and (b) logo invisibility rate among successful injections in the poison-based attack (this work) and Silent Branding~\cite{jang2025silentbranding}.}
  % \label{fig:poison_ablation}
\end{figure}

To better understand whether the poisoned image generation model itself can reliably inject hidden logos before being edited by any downstream models, we conduct ablation experiments on the poisoned model alone.

% We analyze the poisoned model, without including any downstream editing model.
We begin with the logo \textit{injection rate}, \ie{} how often the poisoned image generation model successfully embeds a hidden logo in its output image. As shown in \Cref{fig:poison_embed_rate}, our method achieves substantial injection rates across all logos. Among them, ``Chanel'' has the lowest injection rate, likely because its logo contains multiple intersecting lines and is therefore more difficult for the model to learn during fine-tuning.
We next evaluate the \textit{invisibility rate} among successful injections. 
\Cref{fig:poison_model_invisibility} shows that the invisibility rates remain high across all logos, indicating that a large fraction of successful injections are also visually imperceptible. 
These results suggest that the poison-based attack achieves both visual stealthiness and reliable logo injection.

% both goals discussed in \Cref{sec:feasibility}---visual stealthiness and reliable logo injection.

We further compare our method against Silent Branding~\cite{jang2025silentbranding}, since successful hidden logo injection in our setting depends on the fine-tuning strategies introduced in \Cref{subsec:poison-design}. 
This comparison helps isolate the contribution of our design choices.
\Cref{fig:poison_embed_rate} shows that Silent Branding's injection rate is low and nonzero only for ``Apple'' and ``McDonald's.'' 
This is likely because these brands are common and their logos are occasionally rendered by the model.
In addition, the invisibility evaluation in \Cref{fig:poison_model_invisibility} shows that all successful injections from the Silent Branding baseline are visible. Note that ``N/A'' denotes cases with zero successful injections, and 0\,\% denotes cases in which every successful injection is noticeable.

\begin{figure*}[t]
  \centering
  \includegraphics[width=\linewidth]{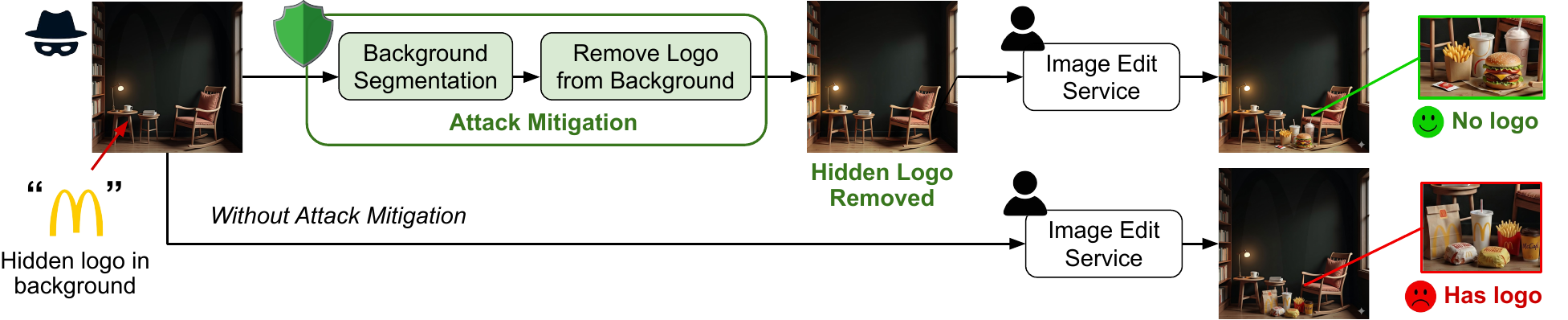}
  \caption{Workflow of attack mitigation mechanism. With the mitigation model, the hidden logo is removed and the editing model does not render the logo in the output image. }
  \label{fig:defense_workflow}
\end{figure*}

\subsection{Discussion} \label{subsec:poison-discussion}

The poison-based attack demonstrates that hidden branding injection can also be realized as a supply-chain attack.
Unlike the phishing-based scenario, the attacker no longer controls the generation service, the user prompt, or the returned image after deployment. 
Instead, the attacker poisons a text-to-image model in advance and distributes it as a seemingly benign checkpoint. 
Once the victim uses this compromised model for local generation, the produced image may already contain a weak hidden logo.
When the image is later edited by a downstream model, that hidden hint can be amplified into a visible branded object. 
This makes the threat particularly concerning for users who rely on local deployment for safety, because the compromise is embedded in the model weights and can silently propagate across later editing stages.

This scenario is related to prior work on poisoning and backdoor attacks against diffusion models, but differs in an important way. 
Existing studies show that poisoned text-to-image models can learn attacker-chosen patterns and be triggered by inputs~\cite{Nightshade,sun2026attacksapproximatecachestexttoimage,jang2025silentbranding}.
Silent Branding~\cite{jang2025silentbranding} injects logos by poisoning the training dataset so that the resulting model automatically generates the corresponding logos; Nightshade~\cite{Nightshade} further shows that even a small amount of poisoned data can control an entire category of generations.
CachePoison~\cite{sun2026attacksapproximatecachestexttoimage} injects logos by exploiting the approximate cache deployed in text-to-image serving systems. 
These methods only focus on a single model scenario. 
In comparison, our poison-based attack highlights a different risk: a compromised upstream generator can plant hidden information that survives transfer to a separate downstream editor, revealing a cross-model supply-chain threat that has not been studied.

% \sihang{Meng, can you take a look?}
% \mx{edited a bit}
This poison-based attack is also related to model-based watermarking approaches.
For example, Tree-Ring Watermarks \cite{wen2023treerings} fingerprints diffusion output by embedding special patterns into the initial noise of diffusion and Stable Signature \cite{Fernandez_2023_ICCV} fine-tunes the model's latent decoder to embed signatures into outputs.
Like post-generation watermarking (discussed in \Cref{subsec:phishing-discussion}), these watermarking methods target provenance or verification, a completely different scenario from this work.

\begin{figure}[t]
  \centering
  \includegraphics[width=\linewidth]{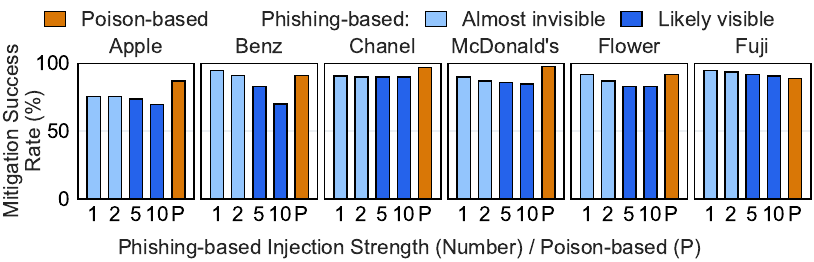}
  \caption{Mitigation success rates for phishing-based attack.}
  \label{fig:defense_success_rate}
\end{figure}

\begin{figure*}[t]
  \centering
  \includegraphics[width=1\linewidth]{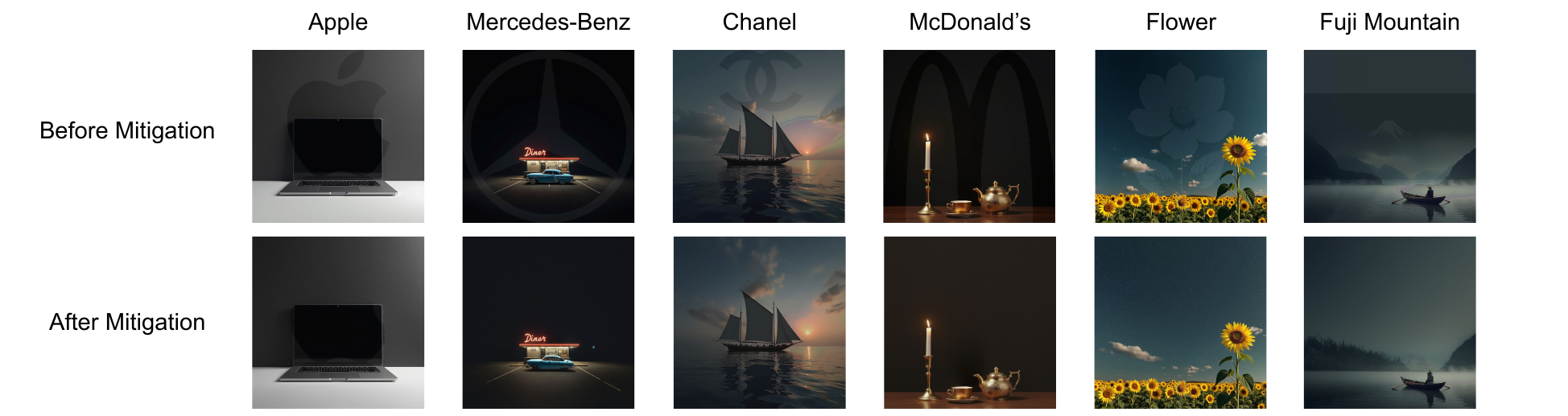}
  \caption{Example images from the HQ-Edit dataset under the phishing-based attack, including images before and after mitigation. The logo injection strength is set to 10 for clarity. }
  \label{fig:defense_examples}
\end{figure*}

\section{Attack Mitigation}

In this section, we introduce our attack mitigation solution and evaluate its effectiveness.

\subsection{Mitigation Design}
% Although these attacks are effective, there are still techniques that can mitigate the harmness.

% \sihang{Meng, can you check this? }
% \mx{I think it makes sense logically. Ideally we might need to run an experiment to quantify the cost, (i.e., how much it cost for an LLM to "detect" whether there is any logo embedded) but if we don't have much time before the submission, it is fine.}
% Because the attacker may introduce any arbitrary logo by injecting them through a phishing service or a poisoned model and we cannot know the shape of the possible logo in advance, actively detecting logos before image editing is futile. 
Because the attacker may inject arbitrary logos or patterns, the mitigation mechanism has no prior knowledge about the injected payload. 
Actively detecting arbitrary hidden logos in the image and determining whether they pose a threat using a multi-modal model can incur a significant overhead relative to the cost of image editing. % as it would require a vision-capable multi-modal model to perform detection.
For reference, as of May 2026, the multimodal model GPT-5.4 costs \$2.5 per 1M tokens, whereas the editing model we evaluate, GPT Image 1.5, costs \$8 per 1M tokens~\cite{openai-pricing}.

% This is difficult and expensive to identify the injected payload correctly and only edit the corresponding part.}
Therefore, we aim to take a lightweight approach. 
As demonstrated in \Cref{subsec:motivation-background}, the attack is most effective when the hidden pattern is placed in a low-entropy region such as a smooth or nearly uniform background. On top of that, the attacker improves stealthiness by making the embedded logo visually similar to the surrounding background, so that it becomes difficult for users to notice while being recognizable by a later editing model.
These observations suggest a more practical mitigation strategy, that is to reduce the attack success rate by breaking these conditions.
If a preprocessing step can disrupt either the low-entropy region or the similarity between logo and the background, the downstream editing model is less likely to preserve and amplify the hidden logos.

\Cref{fig:defense_workflow} illustrates our mitigation workflow. 
Because the attack is most effective when the logo is hidden in a low-entropy background, we first use a segmentation model to identify simple, smooth background regions that are likely to be exploited by the attack. We then use the resulting mask to guide a lightweight inpainting model that regenerates these low-entropy regions.
As discussed earlier, because the attacker can introduce any logo, the inpainting model in our mitigation pipeline uses a generic prompt for image cleaning and restyling without mentioning any logo-specific information. At the same time, it preserves the intended high-contrast visual patterns in the image.
After this lightweight cleaning step, the hidden logo no longer exists and the image is forwarded into the edit service as requested by the user.

\subsection{Experiment Setup}
\textbf{System Platform:} We take the same system as in \Cref{subsec:phishing-setup}.

\textbf{Models and Dataset.} We use SAM2 Large (224M parameters) \cite{ravi2024sam2} as the segmentation model to detect background regions, and a lightweight FLUX.1 Fill (12B parameters)~\cite{flux2024} model to remove the hidden logo.
Note that both models involved in the mitigation process are relatively lightweight, as their scales are much smaller than hundred-billion-parameter commercial image editing models \cite{articsledge2026modelparameters} (such as Gemini).
For the phishing-based and poison-based attacks, we follow the same attack setups as in \Cref{subsec:phishing-setup,subsec:poison-setup}, respectively.

\textbf{Evaluation Metrics.} We evaluate the mitigation effectiveness with the following metrics:

\begin{itemize}[leftmargin=*]
\item \textbf{Success rate:} We also use the Qwen3-VL 30B \cite{yang2025qwen3technicalreport} to detect whether the image after mitigating contains a logo. If an image does not contain such a logo after mitigation, we consider it as successful.
\item \textbf{CLIP Score:} We adopt CLIP to evaluate the alignment between image and generation prompt \cite{hessel-etal-2021-clipscore}, like our method in \Cref{subsec:phishing-setup}. % A higher CLIP score means the image is more similar to user's expectation.
\end{itemize} 

\subsection{Mitigation Evaluation}

\subsubsection{Mitigation Success Rate}
We first evaluate the success rate of our mitigation mechanism against both the phishing-based and poison-based attacks. 
% To better quantify the mitigation effectiveness, we evaluate our mitigation mechanism on phishing-based attack as its logo injection strength can be controlled, and 
For the phishing-based attack, we select four levels of injection strength, the same methodology as in \Cref{subsubsec:phishing-embed-level-success-rate}.
\Cref{fig:defense_success_rate} shows that our mitigation is effective across all six logos.
In the phishing-based scenario, when the hidden logo remains nearly invisible before editing, the mitigation success rate reaches 87.4\,\% at the default injection strength level of 2. 
When the injected logo becomes more visible, the success rate drops slightly (81.4\,\% at level 10). 
However, at higher injection strengths, users are more likely to notice the logo, which reduces the potential risk.
In the poison-based scenario, the mitigation success rate reaches 92.3\,\%.

Among the six logos, ``Apple'' has the lowest mitigation rate. This is likely because the model can render this logo even without the hidden logo because it is a common brand. This is consistent with our earlier results that ``Apple'' has the highest rendering rate in the no-attack setting, as illustrated in \Cref{fig:attack_success_rate_phishing} and \Cref{fig:attack_success_rate_poison}.  

% This result confirms that finding the low-entropy region and smoothing it effectively suppresses hidden logo injections before editing.

% Nevertheless, the overall trend remains encouraging: even for stronger attacks, the mitigation still prevents the downstream editor from rendering the attacker-controlled logo in most cases. 

\begin{figure}[t]
  \centering
  \includegraphics[width=\linewidth]{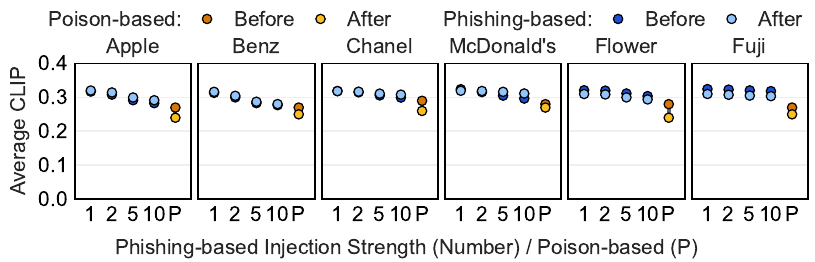}
  \caption{CLIP scores before and after applying mitigation to the phishing-based attack.}
  \label{fig:defense_clip}
\end{figure}

% \begin{figure}[t]
%   \centering
%   \includegraphics[width=\linewidth]{fig/evaluation/defense-ssim.pdf}
%   \caption{SSIM scores that measure the structural similarity before and after applying the mitigation.}
%   \label{fig:defense_ssim}
% \end{figure}

% We further compare the clip score before and after defense mechanism.
\subsubsection{Mitigation Quality}
\Cref{fig:defense_clip} measures whether the mitigation affects the semantic alignment between the image and the prompt.
The average CLIP scores before and after mitigation remain close across all six logos. 
When the injection strength is high in the phishing-based scenario, the score increases after mitigation because highly visible logos can degrade the visual quality.
In the poison-based scenario, the quality degradation in the customized ``Flower'' logo is higher because of its more complex patterns. 
This result suggests that the mitigation primarily targets the hidden logos, and removing them does not noticeably degrade the image semantics requested by the user.

\Cref{fig:defense_examples} demonstrates examples from the phishing-based attack.
We use an injection strength of 10 to better visualize the logos. 
% Before mitigation, the images contain hidden logos in the background which can be recognized by the downstream editing model.
The background patterns injected by the attacker are completely removed after mitigation, while the image preserves the original semantics.

% \subsubsection{Performance Overhead}
% The mitigation involves a segmentation model and an inpainting model. 
% We measure the latency of both steps on our local testbed. 
% On average, segmentation takes \SI{0.14}{\second} and inpainting takes \SI{22.9}{\second}. 
% As the mitigation mechanism will be deployed on the editing model side, its overhead is relative low (the inpainting model has only 12B parameters), compared to that of hundred-billion-parameter commercial image editing models \cite{articsledge2026modelparameters} (such as Gemini). 
% \mx{Since this is an overhead number, I think it is better to show the original time it takes to edit the image, without segmentation and inpainting. (say, 80 s), and then we can argue that the overhead of about 23 s is not too much, and the reason why the overhead is small is because we use a 12B model for inpainting.}

% These examples illustrate the main advantage of the defense: it disrupts the low-entropy carrier region where the hidden logo resides, instead of aggressively rewriting the whole image.

\subsection{Discussion}
Our mitigation solution uses a lightweight preprocessing step on images before they are sent to a downstream editing model. Instead of trying to identify a specific logo, the mitigation targets the conditions that make the attack effective: it detects low-entropy background regions and then regenerates them with a lightweight inpainting model.
Therefore, this mechanism is model-agnostic. 
% In particular, it detects the low-entropy background region and then regenerates that region with a lightweight inpainting model. 
% As a result, the mitigation breaks the logo pattern, reducing the chance that the editor reconstructs the attacker-chosen brand while preserving the overall image semantics.

Prior studies have also explored removing such inconspicuous patterns. One direction uses compression to remove fragile signals~\cite{SHIELD,Feature-Distillation,zhao2024invisible}. Our mitigation mechanism differs from these strategies, as our attack relies on a semantically meaningful but spatially low-strength logo. 
Moreover, our mitigation mechanism does not require access to the internals of either the generation model or the final editing model, which makes it practical and generic.

\iffalse
This mitigation mechanism also has two practical advantages. First, it is model-agnostic: it does not require access to the internals of either the generation model or the final editing model. Second, it is lightweight and inexpensive to deploy, because the cleaning step can be performed by a relatively small model before the image is sent to a more expensive editing service.
\fi
% In this way, the mitigation system uses an inexpensive edit model to remove fragile hidden cues and then relies on texture injection to reduce the chance that any remaining signal survives into the final high-quality output.
% \sihang{Also discuss related works, like using compression, and say they fundamentally don't work. }

% \input{sec/8_related_works.tex}
\section{Conclusions}

In this paper, we identify a novel security vulnerability in multi-phase, cross-model image generation workflows, where nearly invisible hints embedded in an initial image can be recognized by downstream editing models and rendered as visible branded content. We show that this threat is practical in both a phishing-based scenario, in which an attacker controls a malicious image generation service, and a poison-based scenario, in which the attacker distributes a poisoned image generation model. In both scenarios, hidden logos can be rendered onto objects in the edited output with high success rates. We also present a mitigation solution that removes such hidden logo patterns before editing.

\newpage
\section{Ethics Considerations}

\iffalse
This work involves no human subjects and does not rely on private or sensitive user data. All experiments involving online image-editing APIs were conducted only through the authors' own accounts, and we did not interfere with other users or public services. 
Our poisoning experiments were restricted to locally deployed open-source models under our control, so the modified models did not affect external users or third-party systems. 
In addition, the injected logos were used only as proof-of-concept for demonstrating the security risk; we did not use illegal, hateful, violent, or otherwise harmful content in any experiment. 
% We also do not release any poisoned models or attack-ready artifacts, and we disclose this vulnerability to help the community better understand and mitigate misuse in multi-stage image-generation workflows.
\fi

This work does not involve any human subjects and does not rely on any private or sensitive user data. 
All experiments involving online image editing APIs were conducted only through the authors’ accounts, and we did not interfere with other users, public services, or third-party systems. 
Our poisoning experiments were restricted to locally deployed open-source models under our control, so the modified models did not affect external users or third-party systems. 
The injected logos and patterns were used only as proof-of-concept payloads to demonstrate the security risk; we did not use illegal, hateful, violent, or otherwise harmful content in any experiment.

We recognize that the attack techniques studied in this paper could be abused to inject unwanted content into image generation-and-editing workflows. 
The goal of this work is to characterize the vulnerability and develop defenses, rather than to enable deployment of the attack.
% We will not release poisoned model checkpoints, attack-ready services, or artifacts that directly enable hidden-payload injection. 
Any released code or data will be limited to benign evaluation and mitigation components, without including any harmful data. 
% We also plan to notify relevant model and service providers about the vulnerability and our findings before public release, following responsible disclosure practices.

\bibliographystyle{IEEEtran}
\bibliography{bib/sec,bib/ml,bib/sys,bib/misc}

\end{document}